\documentstyle[a4,12pt]{article}
\input{psfig.sty}

\begin{document}

\setcounter{section}{0}
\setcounter{equation}{0}
\def\theequation{\arabic{section}.\arabic{equation}}

\begin{titlepage}

\baselineskip 24pt

\begin{center}
{\Large {\bf The Dualized Standard Model and its Applications---an Interim
   Report}}

\vspace{.5cm}

\baselineskip 14pt
{\large CHAN Hong-Mo}\\
chanhm\,@\,v2.rl.ac.uk \\
{\it Rutherford Appleton Laboratory,\\
  Chilton, Didcot, Oxon OX11 0QX, United Kingdom}\\
\vspace{.2cm}
{\large TSOU Sheung Tsun}\\
tsou\,@\,maths.ox.ac.uk\\
{\it Mathematical Institute, University of Oxford,\\
  24-29 St. Giles', Oxford OX1 3LB, United Kingdom}
\end{center}

\vspace{.3cm}

\begin{abstract}

Based on a nonabelian generalization of electric-magnetic duality, the
Dualized Standard Model (DSM) suggests a natural explanation for exactly 3 
generations of fermions as the `dual colour' $\widetilde{SU}(3)$ symmetry
broken in a particular manner.  The resulting scheme then offers on the one 
hand a fermion mass hierarchy and a perturbative method for calculating the 
mass and mixing parameters of the Standard Model fermions, and on the other 
testable predictions for new phenomena ranging from rare meson decays to 
ultra-high energy cosmic rays.  Calculations to 1-loop order gives, at 
the cost of adjusting only 3 real parameters, values for the following 
quantities all (except one) in very good agreement with experiment: the 
quark CKM matrix elements $|V_{rs}|$, the lepton CKM matrix elements 
$|U_{rs}|$, and the second generation masses $m_c, m_s, m_\mu$.  This means, 
in particular, that it gives near maximal mixing $U_{\mu3}$ between $\nu_\mu$ 
and $\nu_\tau$ as observed by SuperKamiokande, Kamiokande and Soudan, while
keeping small the corresponding quark angles $V_{cb}, V_{ts}$.  In addition,
the scheme gives (i) rough order-of-magnitude estimates for the masses of 
the lowest generation, (ii) predictions for low energy FCNC effects such as 
$K_L \rightarrow e \mu$, (iii) a possible explanation for the long-standing
puzzle of air showers beyond the GZK cut-off.  All these together, however, 
still represent but a portion of the possible physical consequences derivable
from the DSM scheme the majority of which are yet to be explored.  

\end{abstract}

\end{titlepage}

\clearpage

The Dualized Standard Model (DSM) is an attempt to answer some of the 
questions left open by the Standard Model, such as the reason for the 
existence of 3 fermion generations and of Higgs fields, and to explain 
the observed values of some of the Standard Model's many parameters.  
In contrast to most attempts with a similar purpose which extend or
modify the SM's theoretical framework, the DSM scheme remains within it, 
only going beyond the SM in recognizing and exploiting a nonabelian 
generalization of electric-magnetic duality discovered in Yang-Mills 
theory a couple of years ago.  Thus, it is just by assigning physical 
significance to the newly discovered dual symmetries which are claimed 
to be inherent already in the SM gauge theory that the DSM obtains its 
new physical results. 

In this paper we shall briefly review its basic assumptions, outline 
its main features and summarize the results so far obtained from its 
phenomenological applications.\footnote{Readers preferring an even shorter
summary is referred to our report with J. Bordes \cite{Vancouver} given 
at ICHEP'98 Vancouver.}  This is merely an interim report in that only a 
portion of the new potentials opened up by the scheme has been explored, 
and even in that portion there are indications that some of the assumptions 
and approximations made could (or perhaps even need to) be both tightened 
and refined.  Besides, there are yet unanswered questions of consistency 
both within the scheme itself and of the scheme with nature.  However, 
even with these limitations, the results so far obtained are, we believe, 
already encouraging enough to be interesting.

As mentioned above, the DSM scheme is based on a generalization \cite{dualsymm}
of the familiar electric-magnetic duality of electromagnetism to nonabelian 
Yang-Mills theory, the full development of which requires a fair amount 
of theoretical apparatus formulated in loop space, and is therefore 
beyond the scope of the present review aimed mainly at phenomenological 
applications.  Fortunately, for our present purpose, very little detail
of the theory is required, which we shall have no difficulty later briefly 
to supply.  For the reader, however, who is interested in the theoretical
bases, we have written a companion paper to the present one giving a short
review of the steps involved in deriving them \cite{DSMrth98}.

\section{Nonabelian Duality and Basic Ingredients of DSM}

To illustrate how nonabelian duality enters in the Standard Model, we 
begin with a short resum\'e of (abelian) duality in electromagnetism.  
The Maxwell theory has long been known to possess a symmetry under the 
interchange of electricity and magnetism:
\begin{equation}
F_{\mu\nu}(x) \longleftrightarrow {}^*\!F_{\mu\nu}(x),
\label{dualsymm}
\end{equation}
where the *-operation (Hodge star):
\begin{equation}
{}^*\!F_{\mu\nu}(x) = -\frac{1}{2} \epsilon_{\mu\nu\rho\sigma} 
F^{\rho\sigma}(x),
\label{dualtransf}
\end{equation}
we may call the abelian dual transform.  Electric charges are sources of 
the field $F$ but monopoles of the field ${}^*\!F$, while magnetic charges are 
monopoles of $F$ but sources of ${}^*\!F$, where the strength of the quantized 
electric charges $e$ is related to that of the also quantized magnetic 
charges $\tilde{e}$ by the famous Dirac condition:
\begin{equation}
e \tilde{e} = 2 \pi.
\label{diraccond}
\end{equation}
At any point $x$ in space-time free (locally) of electric and magnetic 
charges, both $F_{\mu\nu}(x)$ and its dual ${}^*\!F_{\mu\nu}(x)$ are, by 
virtue of the Maxwell equations, `gauge fields' derivable from potentials, 
thus:
\begin{equation}
F_{\mu\nu}(x) = \partial_\nu A_\mu(x) - \partial_\mu A_\nu(x),
\label{fina}
\end{equation}
\begin{equation}
{}^*\!F_{\mu\nu}(x) = \partial_\nu \tilde{A}_\mu(x) - \partial_\mu 
\tilde{A}_\nu(x).
\label{dftinda}
\end{equation}
It follows then that the theory is invariant under the 2 independent gauge
transformations:
\begin{equation}
A_\mu(x) \longrightarrow A_\mu(x) + \partial_\mu \alpha(x),
\label{gaugetransf}
\end{equation}
\begin{equation}
\tilde{A}_\mu(x) \longrightarrow \tilde{A}_\mu(x) + \partial_\mu
   \tilde{\alpha}(x).
\label{dgaugetransf}
\end{equation}
In other words, the theory is invariant under the doubled gauge symmetry 
$U(1) \times \tilde{U}(1)$, although obviously, given (\ref{dualtransf}), 
the dual fields $F$ and ${}^*\!F$ represent just the same physical degrees of 
freedom.  Notice that this `doubled' gauge symmetry is inherent in the 
Maxwell theory itself and not an additional assertion imposed from outside.  
The only reason we are less familiar with, and have not made much used of,
the dual gauge symmetry $\tilde{U}(1)$, which theoretically is on an equal 
footing with the other gauge symmetry $U(1)$, is just that in the physical 
world we have observed electric charges but not so far (apparently) their 
magnetic counterparts.  

Nonabelian Yang-Mills theory is not symmetric under the *-operation of
(\ref{dualtransf}) \cite{Guyang} so that it was not known for some time
whether the dual symmetry of electromagnetism generalizes to Yang-Mills theory.
But it has now been shown in \cite{dualsymm} that there is a generalized 
nonabelian dual transform \ $\tilde{\ }$\ , reducing to * in the abelian case, 
under which Yang-Mills theory is symmetric.  Unfortunately, the explicit 
formula for the generalized transform\ $\tilde{\ }$\   is known at present 
only in the language of loop space \cite{Polyakov,book} and is for 
that reason a little cumbersome.  However, for our present discussion, we 
need only to note the consequences of the symmetry it implies, as follows.

Dual to a Yang-Mills field $F_{\mu\nu}$ derivable from a potential $A_\mu$:
\begin{equation}
F_{\mu\nu}(x) = \partial_\nu A_\mu(x) -\partial_\mu A_\nu(x)
   + ig\, [A_\mu(x), A_\nu(x)],
\label{FinA}
\end{equation}
there is a field $\tilde{F}_{\mu\nu}$ (the relation of which to $F_{\mu\nu}$ 
is known though complicated) which is also derivable from a potential, 
$\tilde{A}_\mu$:
\begin{equation}
\tilde{F}_{\mu\nu}(x) = \partial_\nu \tilde{A}_\mu(x) - \partial_\mu 
   \tilde{A}_\nu(x) + i \tilde{g}\, [\tilde{A}_\mu(x), \tilde{A}_\nu(x)],
\label{dFindA}
\end{equation}
where the coupling strengths $g$ and $\tilde{g}$ are related by a generalized
Dirac condition \cite{dualcomm}:
\begin{equation}
g \tilde{g} = 4 \pi.
\label{Diraccond}
\end{equation}
`Colour' (electric) charges appear as sources of the field $F$ but monopoles 
of the dual field $\tilde{F}$, while `dual colour' (magnetic) charges appear 
as monopoles of $F$ but sources of $\tilde{F}$ \cite{dualsymm,dualsym}.  
The symmetry claimed in \cite{dualsymm} then means that Yang-Mills theory 
is invariant under the 2 independent gauge transformations:
\begin{equation}
A_\mu(x) \longrightarrow A_\mu(x) + \partial_\mu \Lambda(x)
   + ig\, [\Lambda(x), A_\mu(x)],
\label{Gaugetransf}
\end{equation}
and
\begin{equation}
\tilde{A}_\mu(x) \longrightarrow \tilde{A}_\mu(x) + \partial_\mu 
   \tilde{\Lambda}(x) + i \tilde{g}\, [\tilde{\Lambda}(x), \tilde{A}_\mu(x)],
\label{dGaugetransf}
\end{equation}
giving thus to a theory with structure group $G$ a `doubled' gauge symmetry 
$G \times \tilde{G}$, in close analogy to the situation in electromagnetism.  
This doubling of the gauge symmetry, as in the abelian case, is claimed
to be an inherent property of the gauge theory, not an additional input.
Again, there is no doubling in the physical degrees of freedom.  

Given this nonabelian duality let us now examine its implications in 
the Standard Model with the structure group $SU(3) \times SU(2) \times 
U(1)$.\footnote{Here, for ease of presentation, we ignore the subtle 
differences between gauge groups with the same algebra, although of 
course global properties of the gauge group are of primary importance 
in considerations of monopole charges.  (See e.g. \cite{book}.)}  
From the discussion in the preceding paragraph, it follows that the 
theory will have, in addition to the familiar gauge symmetry $SU(3)
\times SU(2) \times U(1)$, also the dual gauge symmetry $\widetilde{SU}(3) 
\times \widetilde{SU}(2) \times \tilde{U}(1)$.  This dual gauge symmetry 
being, according to our observations above, an inherent property of the 
gauge theory, what we call the Dualized Standard Model \cite{dualcons} 
which utilizes this dual gauge symmetry is, as at present understood, 
no different in principle as a gauge theory from the Standard Model itself.  
As physical schemes, however, they differ, but only in the DSM scheme's 
recognition of the existence of and its assignment of physical meanings
to the dual gauge symmetry which is ignored in the usual SM treatment.
In other words, the new DSM results are deduced just by exploring how 
this `pre-existing' dual gauge symmetry is likely to manifest itself 
in the physical world.  The exploration is based on two main assumptions,
namely the identification of 2 ingredients inherent in the theory to 2 
physical objects, as we shall now explain.

The first assumption concerns the physical interpretation of dual colour.
Indeed, the DSM results so far obtained are all from exploitations of 
only the $\widetilde{SU}(3)$ dual colour symmetry, and it is to this 
symmetry that our considerations in this paper will be restricted.  
Colour symmetry $SU(3)$ being known from experiment to be confined, 
't~Hooft's famous arguments \cite{tHooft} then suggest that the dual 
colour symmetry $\widetilde{SU}(3)$ should be broken.\footnote{That 
the definition of duality given in \cite{dualsymm} is consistent with 
't~Hooft's definition given in \cite{tHooft} is shown in \cite{dualcomm}.}  
Now the idea has long been toyed with that fermion generations may 
be considered as a broken `horizontal' symmetry.  In most schemes, 
this new symmetry will have to be introduced {\it ad hoc}.  Here, 
however, given the fact that because of duality and the arguments of 
't~Hooft, there is inherent in the SM gauge theory already a broken symmetry 
$\widetilde{SU}(3)$, it seems natural to explore the possibility of 
making it into the `horizontal' symmetry of generations.  The idea
is made particularly attractive by the fact that recent LEP experiments
have determined the number of generations of light neutrinos to be 3
to a high accuracy \cite{Lep3}.  Hence, the first basic assumption one
makes in the DSM scheme is that dual colour is to be identified with 
generations.  This means, first, that there will be just 3 generations, 
no more no less, and second, that any particle carrying a generation 
index will carry a dual colour charge, implying, as explained above, 
that it is a colour monopole of the Yang-Mills field $F$.  In particular 
a quark will be a colour `dyon' with both colour (electric) and dual 
colour (magnetic) charges.

Now, 't~Hooft's arguments suggest that the dual colour symmetry is broken, 
but they do not tell us explicitly how this breaking will occur.  The 
interesting thing is that within the DSM gauge theory, there are 
quantities which can figure as Higgs fields, and if so identified will 
imply a particular symmetry breaking pattern.  The candidates as Higgs 
fields in the DSM are the frame vectors in internal symmetry space, which
in the case of dual colour is the space of $\widetilde{SU}(3)$.  The 
idea of using frame vectors as dynamical variables is made familiar 
already in the theory of relativity where in the Palatini treatment or 
the Einstein-Cartan-Kibble-Sciama formalism \cite{Heyl} the space-time 
frame vectors or vierbeins are used as dynamical variables.  In gauge 
theory, frame vectors in internal symmetry space are not normally given 
a dynamical role, but it turns out that in the dualized framework they 
seem to acquire some dynamical properties, in being patched, for example, 
in the presence of monopoles \cite{dualsym}.  Moreover, they are 
space-time scalars belonging to the fundamental representations of the 
internal symmetry group, i.e. doublets in electroweak $SU(2)$ and 
triplets in dual colour $\widetilde{SU}(3)$, and have finite lengths 
(as vev's).  They thus seem to have just the right properties to be 
Higgs fields, at least as borne out by the familiar example of the 
Salam-Weinberg breaking of the electroweak theory.  Hence, one makes 
the second basic assumption in the DSM scheme, namely that these frame 
vectors are indeed the physical Higgs fields required for the spontaneously 
broken symmetries.

Identifying dual colour with generations and frame vectors with Higgs
fields are of course assumptions.  It is worth noting, however, that
even if one does not choose to do so, the niches in the form of dual
colour and frame vectors will still exist and manifest themselves
physically in some way which will need to be accounted for in another 
manner.  Opting for the identification, on the other hand, not only 
offers a hope of determining some of the SM parameters but also gives 
a much desired geometric significance both to generations and to
Higgs fields which has been sadly lacking to an otherwise highly 
geometric theory.

Having made these basic assumptions, let us now explore the consequences.
First, making the frame vectors in internal symmetry space into dynamical
variables and identifying them with Higgs fields mean that for dual
colour $\widetilde{SU}(3)$, we introduce 3 triplets of Higgs fields
$\phi^{(a)}_a$, where $(a) = 1, 2, 3$ labels the 3 triplets and $a =
1, 2, 3$ their 3 dual colour components.  Further, the 3 triplets having
equal status, it seems reasonable to require that the action be symmetric 
under their permutations, although the vacuum need not be so symmetric.  
An example of a Higgs potential which breaks both this permutation symmetry 
and also the $\widetilde{SU}(3)$ gauge symmetry completely is as follows 
\cite{dualcons}: 
\begin{equation}
V[\phi] = -\mu \sum_{(a)} |\phi^{(a)}|^2 + \lambda \left\{ \sum_{(a)}
   |\phi^{(a)}|^2 \right\}^2 + \kappa \sum_{(a) \neq (b)} |\bar{\phi}^{(a)}
   .\phi^{(b)}|^2,
\label{Vofphi}
\end{equation}
a vacuum of which can be expressed without loss of generality in terms of 
the Higgs vev's:
\begin{eqnarray}
\phi^{(1)} = \zeta \left(
\begin{array}{c}
 x \\ 0 \\ 0 
\end{array} \right), \,\,\,
\phi^{(2)} = \zeta \left(
\begin{array}{c}
 0 \\ y \\ 0 
\end{array} \right) , \,\,\,
\phi^{(3)} =  \zeta \left(
\begin{array}{c} 
0 \\ 0 \\ z 
\end{array} \right),
\label{vevs}
\end{eqnarray}
with
\begin{equation}
x^2 + y^2 + z^2 = 1,
\label{normvevs}
\end{equation}
and
\begin{equation}
\zeta = \sqrt{\mu/2 \lambda},
\label{zeta}
\end{equation}
$x, y, z$, and $\zeta$ being all real and positive.  Indeed, this vacuum
breaks not just the symmetry $\widetilde{SU}(3)$ but even the larger symmetry 
$\widetilde{SU}(3) \times \tilde{U}(1)$ completely giving rise to 9 massive 
dual gauge bosons.  And of the 18 real components in $\phi^{(a)}_a$, 9 are 
thus eaten up, leaving just 9 (dual colour) Higgs bosons.

Next, we turn to the fermion fields.  In analogy to the electroweak theory,
the left-handed fermion fields are assigned the triplet {\bf 3} and the 
right-handed fermions the singlet {\bf 1} representation in dual colour,
so that one can construct their Yukawa couplings as:
\begin{equation}
\sum_{(a)[b]} Y_{[b]} \bar{\psi}^a_L \phi^{(a)}_a \psi^{[b]}_R,
\label{Yukawa}
\end{equation}
which, by the same reasons as for the Higgs potential, we have made symmetric 
under permutations of the Higgs fields $\phi^{(a)}$.  As a result of this
permutation symmetry, the tree-level fermion mass matrix takes the following
factorized form:
\begin{equation}
m = \zeta \left( \begin{array}{c} x \\ y \\ z \end{array} \right)
   (a, b, c),
\label{mtree}
\end{equation}
where $a, b, c$ are just abbreviations for the Yukawa couplings $Y_{[b]}$.
Written in the hermitized convention of e.g. \cite{Weinberg}, which is 
basically $\sqrt{m m^{\dagger}}$ in terms of the $m$ above and is the 
matrix of relevance to the mass spectrum, it becomes:
\begin{equation}
m = m_T \left( \begin{array}{c} x \\ y \\ z \end{array} \right) (x,y,z),
\label{fermass0}
\end{equation}
where $m_T$ is a mass scale dependent on the fermion type $T = U, D, L, N$, 
i.e. whether $U$-type quarks ($U$), $D$-type quarks ($D$), charged leptons 
($L$) or neutrinos ($N$), but the vector $(x, y, z)$ given in terms of the
Higgs vev's is independent of the fermion type.  

One immediate consequence of such a factorized mass matrix is that at
tree-level there is only one state with mass, represented by the eigenvector
$(x,y,z)$ with eigenvalue $m_T$, the other 2 states having zero eigenvalues.
This we interpret as embryo fermion mass hierarchy, with one generation 
much heavier than the other two generations.  Such an arrangement is not a 
bad first approximation to the experimental situation where the highest 
generation states of $U, D, L$ are all more than one order of magnitude 
heavier than the lower states.  Another immediate consequence is that 
at tree-level the CKM matrix is the identity, since the CKM matrix is the 
matrix giving the relative orientations between the physical states 
of the $U$- and $D$-type quarks, and these orientations at tree-level
are both given in (\ref{fermass0}) by the vector $(x,y,z)$.  Again, 
this is not a bad first approximation to the experimental CKM 
matrix whose off-diagonal elements are at most of order 20 percent.  
Though reasonable as a first approximation, this tree-level description 
is obviously too degenerate to be realistic.  We shall see, however, 
that the degeneracy will be lifted at higher orders where nonzero masses 
for the two lower generations and nonzero off-diagonal elements for the 
CKM matrix will both result from loop corrections.

\setcounter{equation}{0}

\section{Quark Mixing and Light Masses as Loop Corrections}

Radiative corrections to the tree-level mass matrix (\ref{fermass0})
have already been calculated to 1-loop level \cite{ourckm}.  There are
in principle many diagrams to calculate.  There are first the usual
diagrams of the Standard Model with loops of gluons, $\gamma$, $Z$, $W$, 
and electroweak Higgses.  Then there are new diagrams with loops of
dual gluons and dual Higgses.  All these diagrams, however, share
the common feature that they preserve the factorizability of the fermion
mass matrix, namely that after the corrections the mass matrix $m'$ is
of the form:
\begin{equation}
m' = m_T' \left( \begin{array}{c} x' \\ y' \\ z' \end{array} \right)
   (x',y',z').
\label{fermassc}
\end{equation}
The reason is that only those bosons carrying dual colour can affect
the factorizability, and of these the dual gluon couples only to the
left-handed fermion while the dual Higgses have themselves factorizable
couplings.  Indeed, it is believed, though not formally proved, that
factorizability of $m'$ will remain intact to all orders.

\begin{figure}[htb]
\vspace{1cm}
\centerline{\psfig{figure=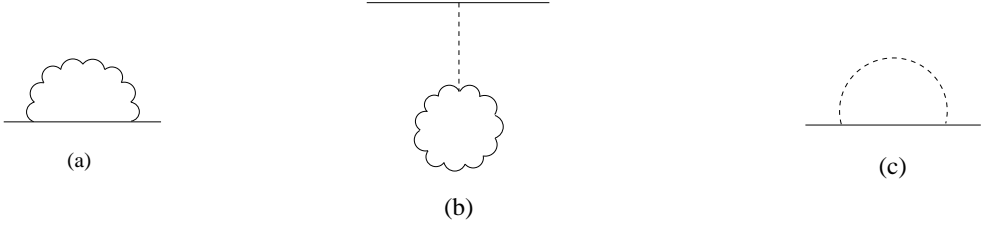,width=0.9\textwidth}}
\vspace{.5cm}
\caption{One loop diagrams rotating the fermion mass matrix.}
\label{oneloopdiag}
\end{figure}

Only a small number of these diagrams, however, need be considered for
the purpose of this paper.  The reason is as follows.  The value of the
normalization factor $m_T'$ in (\ref{fermassc}) is actually not calculable
perturbatively since it receives contributions from diagrams with a dual 
gauge boson loop, and these are proportional to the square of the dual 
gauge coupling $\tilde{g}$ which is large according to the Dirac condition
(\ref{Diraccond}), given the known empirical value of the colour coupling 
$g$.  The normalization $m'_T$ has thus to be treated in any case as an 
empirical parameter.  In other words, the calculation of any diagram 
which affects only this normalization will not add to our present level 
of understanding and might as well be ignored.  What one can calculate 
perturbatively, on the other hand, is the orientation of the vector 
$(x',y',z')$ and this is affected by only a small subset of the diagrams, 
namely those shown in Figure~\ref{oneloopdiag}.  For example, it is clear that
diagrams with only loops of gluons or $W$-bosons will not affect the 
orientation of $(x', y', z')$ since these bosons do not carry dual colour.  
Next, of the remaining 3 diagrams as in Figure~\ref{oneloopdiag}, the 
first 2 can 
give only negligible effects.  This last conclusion is arrived at as follows.  
The diagrams (a) and (b) both give rotations to $(x', y', z')$ but these
are of the order $\tilde{g}^2 m_T^2/\mu_N^2$, where $\mu_N$ represents 
the masses of the dual gauge bosons which are constrained by present 
experimental bounds on flavour-changing neutral current (FCNC) effects to 
be rather large.  The identification of dual colours to generations means
that particles carrying generation indices can interact by exchanging dual
gluons, which would lead to generation-changing or FCNC effects, such as
$K_L \rightarrow e \mu$ decay or an anomalous $K_L - K_S$ mass difference, 
and such effects are strongly bounded by experiment.  An analysis using
the latest data \cite{Cahnrari,ourfcnc} which will be outlined in Section
5 below, shows that a violation to present experimental bounds can be 
avoided only if $\mu_N^2/\tilde{g}^2$ is of order at least a few hundred 
TeV, which means contributions from diagrams (a) and (b), even for the 
top quark, are of at most $10^{-6}$.  Neglecting then contributions of this 
order, the calculation becomes rather simple depending on only the (dual) 
Higgs loop diagram of Figure \ref{oneloopdiag} (c) \cite{ourckm}.

The rotation given to the normalized vector $(x',y',z')$ by the remaining
Higgs loop diagram Figure \ref{oneloopdiag} (c) has been calculated.  It is 
found to depend on the energy scale $\mu$ as follows:
\begin{equation}
\frac{d}{d(\ln \mu^2)} \left( \begin{array}{c} x' \\ y' \\ z' 
   \end{array} \right)
   =  \frac{5}{64 \pi^2} \rho^2 \left( \begin{array}{c}
         \tilde{x}'_1 \\ \tilde{y}'_1 \\ \tilde{z}'_1 \end{array} \right),
\label{runxyz}
\end{equation}
with
\begin{equation}
\tilde{x}'_1 = \frac{x'(x'^2-y'^2)}{x'^2+y'^2} + \frac{x'(x'^2-z'^2)}
   {x'^2+z'^2}, \ \ \ {\rm cyclic},
\label{x1tilde}
\end{equation}
and $\rho^2 = |a|^2 + |b|^2 + |c|^2$ being the Yukawa coupling strength.
As $\mu$ varies the normalized vector traces out a curve on the unit
sphere.  It is
easily seen from (\ref{runxyz}) and (\ref{x1tilde}) that the points 
$(1,0,0)$ and $\frac{1}{\sqrt{3}}(1,1,1)$ are fixed points under scale 
changes, and for decreasing energy, the point $(x',y',z')$ runs away 
from $(1,0,0)$ towards $\frac{1}{\sqrt{3}}(1,1,1)$.  The trajectory
along which $(x',y',z')$ runs depends in principle on the Higgs boson 
masses logarithmically, but because of a `happy accident' that we shall 
make clear later, even this weak dependence can be ignored.  The only
parameters remaining are then just the initial values of the
components of the normalized
vector $(x',y',z')$ identifiable with the Higgs vev's which 
are the same for all fermion types, and the Yukawa coupling strength
$\rho_T$, one for each fermion type $T$.  The information being now 
encoded in the vector $(x',y',z')$ running along a trajectory on the 
unit sphere, the remaining question is just how to extract the fermion 
mass and mixing parameters from this rotating vector.

Let us first consider quarks of the $U$-type and ask what is the physical 
state vector the of $t$ quark in dual colour or generation space.  Now, 
according to (\ref{fermassc}), $(x',y',z')$ is the vector with the single 
nonzero eigenvalue of the loop-corrected fermion mass matrix at the energy 
scale where $(x',y',z')$ is evaluated.  It ought therefore to be 
identified with the state vector of the heaviest generation, except 
that $(x',y',z')$ has to be evaluated at the scale $\mu$ equal to the 
top mass $m_t$.  Inputting then the empirical value of $m_t \sim 176\  
{\rm GeV}$ we run $(x',y',z')$ to the scale $\mu = m_t$.  This gives us 
the physical state vector $|{\bf v}_t \rangle$ and also fixes the 
value of the parameter $m_U'$ at the same scale.  This calculation, 
however, still depends on the starting value $(x,y,z)$ of the rotating 
vector $(x',y',z')$ and also on the Yukawa coupling strength $\rho_U$.

Next, we ask what is the physical state vector of the $c$ quark.  Being 
an independent physical entity, the $c$ quark ought to have a state 
vector orthogonal to $|{\bf v}_t \rangle$.  This means it must have a 
zero eigenvalue of the mass matrix (\ref{fermassc}) evaluated at the 
scale $\mu = m_t$.  But there are two linearly independent such
vectors and we do not as yet know which linear combination of these 
should correspond to the $c$ quark.  Nor can we identify the mass $m_c$
as the zero eigenvalue for this is evaluated at the wrong scale $m_t$.
Let us first extract the mass submatrix in the 2-dimensional subspace 
orthogonal to $|{\bf v}_t \rangle$ and run it down to lower scale.  Being
a submatrix of a rank 1 matrix, it is of rank at most 1 at this lower scale
but need no longer be a zero matrix since the nonzero eigenvector $(x',y',z')$
of the full mass matrix (\ref{fermassc}) has already rotated from the
direction $|{\bf v}_t \rangle$ and can have thus a component in the subspace
orthogonal to $|{\bf v}_t \rangle$ which contains $|{\bf v}_c \rangle$.
For consistency with the way we fixed the vector $|{\bf v}_t \rangle$
before, the procedure to determine $|{\bf v}_c \rangle$ is now clear.
We ought to run the $2 \times 2$ mass submatrix down in energy until
the scale $\mu$ equals the empirical value $m_c$ of the $c$ quark mass.  
Then its only nonzero eigenvector at that scale is to be identified with 
$|{\bf v}_c \rangle$.  This vector is, of course, by definition orthogonal 
to $|{\bf v}_t \rangle$ as it should be.  Its eigenvalue of the mass 
submatrix, however, will not in general be the same as the input value 
of $m_c$.  But our calculation, we recall, still depends on $(x,y,z)$ 
and $\rho_U$, and by adjusting $\rho_U$ we can make the output eigenvalue
of $|{\bf v}_c \rangle$ the same as the input value of $m_c$.  This fixes
$\rho_U$, leaving now only $(x,y,z)$ as parameters.

Once the state vectors of the $t$ and $c$ quarks are determined, then
obviously the state vector of the $u$ quark is also fixed as the vector 
orthogonal to both.  We have thus the whole triad of physical state
vectors of the $U$-type quarks in terms of the 2 remaining parameters in
the normalised vector $(x,y,z)$ representing the Higgs vev's.

The above procedure can now be repeated for the $D$-type quarks and for the
charged leptons to determine the triad of their physical state vectors.  
(The problem for neutrinos is somewhat different as we shall make clear 
later.)  To do so, we shall have to input the empirical masses of the 
two highest generations in each case, namely $b$ and $s$ for the $D$-type
quarks and $\tau$ and $\mu$ for the charged leptons, and then, as before 
for the $U$-type quarks, to adjust the Yukawa coupling strengths $\rho_D$ 
and $\rho_L$ to obtain consistency.  The values of the $\rho$'s determined
in this way need not of course be the same for $D$ and $L$ nor as the 
value obtained before for the $U$-type quarks.  Again the triads so obtained
still depend on $(x,y,z)$, which is the same for all fermion types.

Now the matrix relating the orientations in generation space of the two triads 
of physical state vectors of respectively the $U$-type and $D$-type quarks is 
what is known in the literature as the Cabibbo-Kobayashi-Moskawa (CKM) matrix.  
These orientations being now known, the CKM matrix can be calculated in
terms of the two remaining parameters in $(x,y,z)$.  By adjusting these,
we can then try to fit the empirical CKM matrix.  There are actually 4 
independent degrees of freedom in the CKM matrix, but in our treatment
up to 1-loop order, there is no $CP$-violating phase, the vector $(x',y',z')$
being real.  We are thus left with 3 real quantities to fit with our 
two remaining parameters, which is still nontrivial but has been achieved
rather well.  For the matrix of absolute values $|V_{rs}|$, for $r = u,c,t$
and $s = d,s,b$, we obtained \cite{ourckm} \footnote{This fit was obtained 
using data given in the databook of 1996 \cite{databook}.   For this reason 
the comparison with experiment given in this paper refers for consistency 
also to data from the same source.  Although much of the data have since 
been updated, the changes are not large.  For a parallel fit to the latest 
data and comparison with them, see \cite{phenodsm}, which arrives at a 
similar conclusion to that presented in this paper.}:
\begin{equation}
|V_{rs}| =
\left( \begin{array}{ccc} 0.9755 & 0.2199 & 0.0044 \\
                          0.2195 & 0.9746 & 0.0452 \\
                          0.0143 & 0.0431 & 0.9990 \end{array} \right),
\label{calckmq}
\end{equation}
as compared with the experimental values \cite{databook}:
\begin{equation}
|V_{rs}| =
\left( \begin{array}{lll} 0.9745-0.9757 & 0.219-0.224 & 0.002-0.005 \\
                          0.218-0.224 & 0.9736-0.9750 & 0.036-0.046 \\
                          0.004-0.014 & 0.034-0.046 & 0.9989-0.9993
   \end{array} \right).
\label{expckmq}
\end{equation}

The above fit to the CKM matrix fixes all the parameters in the problem, the 
values of which so obtained show two very intriguing features.  First, the 
vev's of the Higgs fields $(x,y,z)$ turn out to have values very close to 
the high energy fixed point $(1,0,0)$.  Second, and even more intriguingly,
the Yukawa coupling strengths $\rho_T$ turn out to be $T$-independent 
to a surprising accuracy \cite{ourckm}.  Indeed, for fermion masses taken 
at the (geometric) median of experimental values, the fitted values of
$\rho_T$ 
for $T = U, D, L$ differ by only 1.5 parts in a thousand, while varying
the masses within the experimental error bars still give differences of
at most a few percent.  This last `happy accident' seems to indicate 
some hidden symmetry in the scheme, the reason for which we 
have at present only some ideas not yet fully understood.  In practical 
terms, on the other hand, it is a bonus since it simplifies greatly 
not only the calculation but also the presentation of the results.  It 
means, first, that the 3 originally different parameters $\rho_T$ can 
now be treated as just a single parameter $\rho$; second, that all 
fermion types run on the same trajectory at the same speed; and third,
even the originally already weak dependence of the calculation on the 
Higgs boson masses can now be completely ignored.

Next, we turn to the masses of the lowest generation.  Since we know
already the physical state vectors of the lowest generation states for all 
the 3 fermion types $U, D, L$, namely $u, d, e$, we can simply evaluate their
expectation values of the mass matrix $m'$ at any scale.  The actual masses 
of the 3 states are given by these expectation values evaluated at the 
scales equal to the values themselves, and can also be readily calculated.  
However, the calculation of these lowest generation masses is an 
extrapolation on a logarithmic scale and depends also on the
normalization $m'_T$ of the mass matrix $m'$ in (\ref{fermassc}),
whose variation with scale, as we have already 
explained, is not calculable perturbatively.  It is thus expected to be 
far less reliable than the above calculation for the mixing parameters.  
Nevertheless, assuming simply that the normalization $m'_T$ is 
scale-independent, one may hope to get rough order-of-magnitude estimates 
for the lowest generation masses.  The result of such a calculation is 
shown in Table \ref{masstable}.  We notice that the mass of the electron is 
within about an order of magnitude of the experimental value, which we regard 
as reasonable.  As for the quarks, the $d$-quark comes out about right but 
the $u$ quark is too large by nearly two orders of magnitude.  However, 
we should note that light quark masses are notoriously difficult to define,
being sensistive to nonperturbative QCD corrections below around 2 GeV.
Moreover, the $u$ and $d$ masses given in the table are defined each at the 
scale equal to its value whereas the quoted experimental values are defined 
at the scale of 1 GeV.  The two sets of values are therefore not directly 
comparable.  Indeed, if the expectation value in the $u$-state of the running 
mass matrix $m'$ is taken at 1 GeV, a value of order only 1 MeV is obtained,
although it is also unclear whether this should be compared with
the empirical value quoted.  In all cases, at least, the masses are
hierarchical as they should be.
  
\begin{table}
\begin{eqnarray*}
\begin{array}{||c|c|c||}
\hline \hline
& Calculation & Experiment \\
\hline
m_c & 1.327 {\rm GeV} & 1.0-1.6 {\rm GeV} \\
m_s & 0.173 {\rm GeV} & 100-300 {\rm MeV} \\
m_\mu & 0.106 {\rm GeV} & 105.7 {\rm MeV} \\
m_u & 209 {\rm MeV} & 2-8 {\rm MeV} \\
m_d & 15 {\rm MeV} & 5-15 {\rm MeV} \\
m_e & 6 {\rm MeV} & 0.511 {\rm MeV} \\
m_{\nu_1} & 10^{-15} {\rm eV} & < 10 {\rm eV} \\
B & 400 {\rm TeV} & ? \\
\hline \hline
\end{array}
\end{eqnarray*}
\caption{Predicted fermion masses compared with experiment.  Notice, however,
that for the $u$- and $d$-quarks, the calulated masses are defined each at 
the scale equal to its value, and are not directly comparable to the quoted
experimental values defined at the scale of 1 GeV.}
\label{masstable}
\end{table}

\setcounter{equation}{0}

\section{Neutrino Oscillations}

Next, we turn our attention to neutrinos.  The case for neutrino 
oscillations has recently been much strengthened by the atmospheric 
neutrino data from SuperKamiokande \cite{superk} confirming earlier 
results of the last decade \cite{kamioka}--\cite{soudan}.  These have 
not only given convincing evidence for the phenomena, but have even 
provided quite restrictive bounds on the relevant parameters which 
would be a challenge for theoreticians to explain.  

If the basic idea in the DSM scheme of identifying generations with 
dual colour is adhered to, then there will be 3 and only 3 generations 
of neutrinos as for any other fermion type.  In principle then, there 
is nothing to stop us applying the same procedure as that applied above 
to quarks and charged leptons to determine also the masses and physical 
state vectors of neutrinos.  Indeed, in strict adherence to the scheme, 
it would be incumbent upon us to do so.  Since the Higgs vev's $(x,y,z)$ 
are already known, all we need to do so would be to input the (Dirac) 
masses of the two heaviest neutrinos as we did for quarks and charged 
leptons.  In fact, since the Yukawa couplings $\rho$ turned out to 
be so close for all the other fermion types, it seems reasonable 
to assume the same value also for neutrinos.  We shall need then 
to input only one (Dirac) mass.  However, neutrinos differ from
the other fermions in that they can also have Majorana masses, and it
is from these together with their Dirac masses that one obtains their 
physical masses via the well-known seesaw mechanism \cite{Seesaw} by
diagonalizing the matrix:
\begin{equation}
{\bf M}_r = \left( \begin{array}{cc} 0 & M_r \\
                                     M_r & B    \end{array} \right),
\label{seesaw}
\end{equation}
where it turns out that in the DSM scheme as at present understood, all 
3 generations of neutrinos will have to have the same Majorana mass $B$
for consistency \cite{ournuos}.  With then $B$ as an extra parameter, 
we need to input two masses to perform the proposed calculation.

Information on the (physical) mass of the heaviest neutrino, usually 
denoted by $m_3$, is obtained from the muon anomaly in atmospheric 
neutrinos.  From  \cite{kamioka,superk}, we have an estimate of the 
difference between the physical masses of the two heaviest neutrinos.  
Since in the DSM scheme, masses are supposed to be hierarchical, 
meaning $m_3 \gg m_2 \gg m_1$, we can put the mass itself equal to the
difference $m_3^2 \sim 
10^{-3}-10^{-2}\ {\rm eV}^2$, which we can take as one input, leaving 
thus just one more mass to be determined.  

To do so, we draw on the information from solar neutrino data.  There
one has 2 estimates for the (physical) mass of the second generation
neutrino, again taking the masses to be hierarchical.  From the LWO
solution \cite{LWOfit}, one has $m_2^2 \sim 10^{-10}\ {\rm eV}^2$, and
from the MSW solution \cite{MSWfit} $m_2^2 \sim 10^{-5}\ {\rm eV}^2$.  
With either of these as input, one has in principle enough information 
to determine the Dirac mass $M_3$ and hence complete the DSM calculation 
of the leptonic CKM matrix.

It turns out, however, that inputting the MSW estimate for $m_2$ and the 
estimate of \cite{kamioka,superk} for $m_3$, one obtains no sensible DSM 
solution for neutrinos.  The reason is that in the DSM scheme, lower
generation masses come only as a `leakage' from the mass of the highest 
generation and this leakage mechanism does not easily admit a ratio 
$m_2/m_3$ as large as that wanted by the MSW solution.  One concludes 
thus that the DSM scheme, as at present understood, disfavours the MSW
solution to the solar neutrino problem.\footnote{It is interesting to 
note in this context that the latest SuperKamiokande data on day-night
variations and flux reported at the Vancouver ICHEP'98 conference and at the
APS meeting (DPF) at UCLA
in fact also
favour the LWO over the MSW solution for solar neutrinos 
\cite{Vagins,superk}.}

On the other hand, inputting the LWO estimate for $m_2^2 \sim
10^{-10}\  
{\rm eV}^2$ and the estimate of $m_3^2 \sim 10^{-3}- 10^{-2}\ {\rm eV}^2$
from \cite{kamioka,superk}, a solution is readily found.  The state vectors 
of the neutrinos so determined then allow one immediately to calculate the 
leptonic CKM matrix.  For $m_3^2 = 10^{-3} {\rm eV}^2$, one 
obtains \cite{ournuos}:
\begin{equation}
\left( \begin{array}{ccc} U_{e1} & U_{e2} & U_{e3} \\
                          U_{\mu1} & U_{\mu2} & U_{\mu3} \\
                          U_{\tau1} & U_{\tau2} & U_{\tau3} 
       \end{array} \right)
= \left( \begin{array}{ccc} 0.97 & 0.24 & 0.07 \\
                            0.22 & 0.71 & 0.66 \\
                            0.11 & 0.66 & 0.74    \end{array} \right),
\label{calckml}
\end{equation}
all elements being real at the 1-loop level that we are working.  The
result is insensitive to the actual values (in the above range) of $m_2$ 
and $m_3$ used.  Notice that apart from inputting the rough values of 
$m_2$ and $m_3$ from experiment in the way explained, the calculation 
involves no adjustment of parameters which have all been fixed by our 
earlier calculation of the quark CKM matrix \cite{ourckm}.

\begin{figure}[htb]
\vspace{-3cm}
\centerline{\psfig{figure=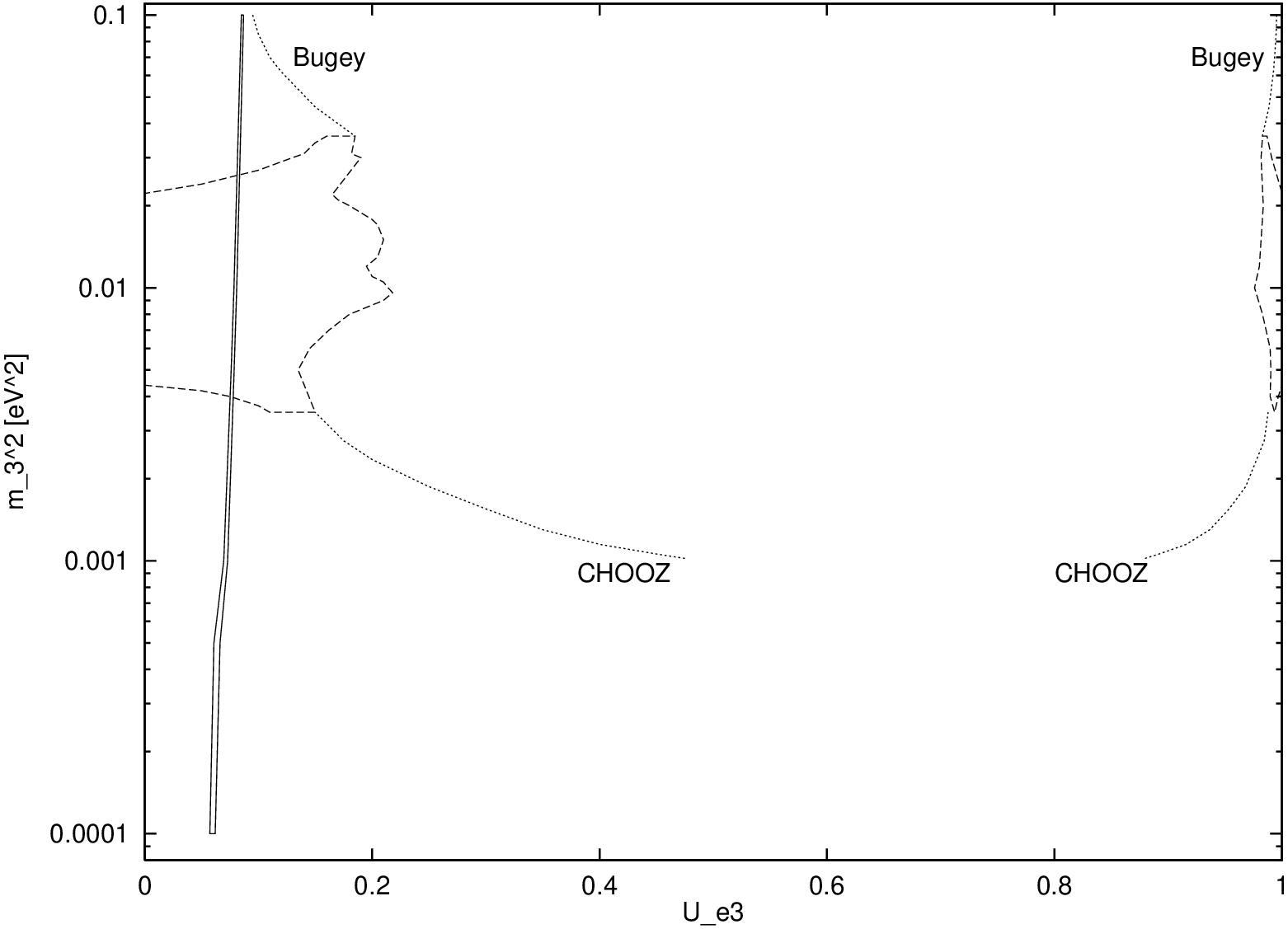,width=0.75\textwidth}}
\caption{90 \% CL limits on the CKM element $U_{e 3}$ compared with
the result of our calculation.}
\label{comp1}
\end{figure}

\begin{figure}[htb]
\vspace{-3cm}
\centerline{\psfig{figure=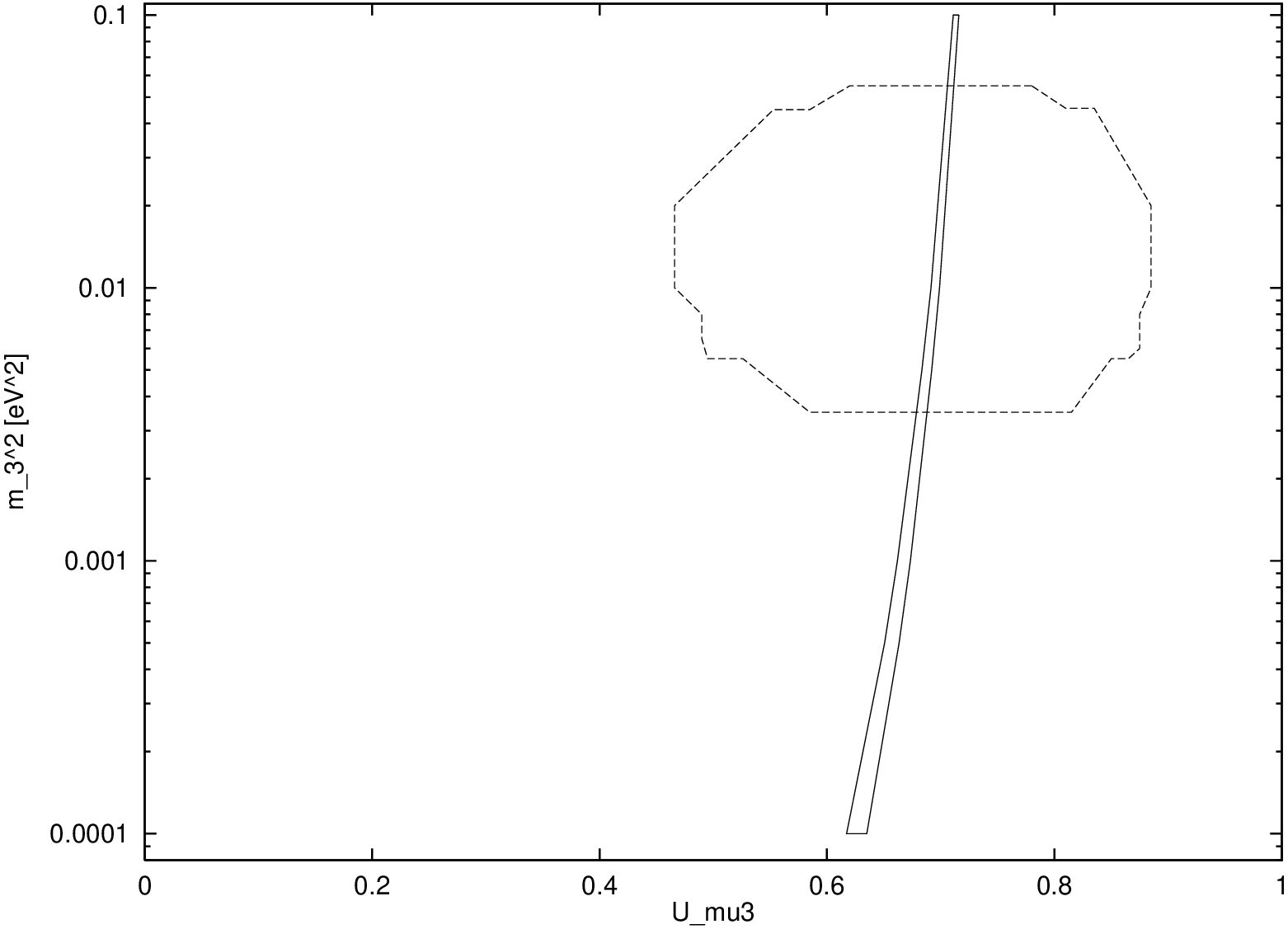,width=0.75\textwidth}}
\caption{90 \% CL limits on the CKM element $U_{\mu 3}$ compared with
the result of our calculation.}
\label{comp2}
\end{figure}

The result (\ref{calckml}) should be compared with the leptonic CKM matrix
extracted from experiment by, for example, 
\cite{Giunkimno}\footnote{This
analysis \cite{Giunkimno} was done with the Kamiokande not with the 
SuperKamiokande data, for which no parallel analysis as far as we know 
has yet been done.  The result, however, is expected to be similar, with
a somewhat tighter bound but roughly the same central value for the 
$\mu 3$ element, but a looser bound on the $e 3$ element.  For a comparison
with the latest data from SuperKamiokande, see \cite{phenodsm}.} 
where the bounds 
on $U_{\mu3}$ comes mainly from atmospheric neutrino data and the bounds on 
$U_{e3}$ comes mainly from reactor data such as \cite{chooz} and the 
estimate for $U_{e2}$ comes from the solar neutrino data as interpreted 
by either the large angle MSW \cite{MSWfit} or the LWO \cite{LWOfit} 
scenario:
\begin{equation}
   = \left( \begin{array}{ccc} \ast & 0.4-0.7 & 0.0-0.2 \\
                               \ast & \ast & 0.5-0.8 \\
                               \ast & \ast & \ast
            \end{array} \right).
\label{expckml}
\end{equation}
Since we are ignoring for the present the CP-violating phase, there 
are only 3 independent elements of the matrix we need consider.  We 
notice that the theoretical predictions for both the angles $U_{\mu 3}$ 
and $U_{e 3}$ fall neatly into the middle of the experimental range.  
A more detailed comparison of our prediction with experiment based 
again on the analysis in \cite{Giunkimno} is shown in Figures \ref{comp1}
and \ref{comp2} for a range of $m_3$ values and for $m_2$  within the 
range allowed by \cite{LWOfit}.  One sees that the agreement is 
consistently good.  
 

The prediction, however, for the other angle $U_{e 2}$ relevant to solar 
neutrinos does not score so well, being about a factor 2 too small and 
lying some way outside the experimental limits.  We shall see later the 
reason why this element is particularly hard for the DSM scheme to get 
correct.

In addition to the mixing angles, the calculation gives predictions also
for the masses of the lightest and the `right-handed' neutrinos, namely
$m_{\nu_1}$ and $B$, as listed in Table \ref{masstable}.  The present 
experimental bound on $m_{\nu_1}$ is too weak to be a test.  As for 
$B$, there is no direct information.  However, given $B$, one can 
estimate within the scheme a value for the life-time of neutrinoless 
double beta decays.  One notes that the value of $B$ we obtained is 
several orders of magnitude lower than that usually given, say for 
example, from grand unified theory models.  The reason is that one 
usually assumes for the heaviest neutrino a Dirac mass similar in value 
to the mass of the $\tau$ or $t$, i.e. of order GeV or higher, whereas the 
value we obtain above by fitting $m_3$ and $m_2$ in the DSM scheme gives a
value only of order MeV (and $B$ is proportional to the square of this
estimate).  Such a big difference between the (Dirac) masses of the 
charged leptons and neutrinos need not be a worry since the same is 
known already to occur between the $U$- and $D$-type quarks.  But as 
a result of this lower value for $B$, the rate for neutrinoless double 
beta decays predicted here will be much more accessible to experiment.  
In particular, a rough estimate shows that the predicted  $0\nu$ half-life 
for ${}^{76}Ge$ is only about 2--3 orders longer than the present 
experimental limit.

To conclude, one sees that the DSM with no freedom left after fitting 
the quark CKM matrix, reproduces quite well the general features of
neutrino oscillations as observed in experiment, and gives in addition
some interesting and in principle testable predictions.  

\setcounter{equation}{0}

\section{Features of Mixing from Differential Geometry}

It is instructive to compare the quark (\ref{expckmq}) and leptonic 
(\ref{expckml}) CKM matrices as now experimentally known.  We note in
particular the following outstanding features: 
\begin{description}
\item{(a)} The corner elements 13 and 31 are much smaller than the 
   others for both quarks and leptons.
\item{(b)} All off-diagonal elements for quarks are much smaller than
   the diagonal elements.
\item{(c)} The 23 element is much smaller for quarks than for leptons.
\end{description}
These will need to be accounted for in any scheme aiming to explain the 
fermion mixing phenomenon.  In the preceding sections we have already 
shown that the DSM scheme is able quantitatively to reproduce these 
features in terms of just a few parameters without explaining why it 
should be so.  What we shall do in this section is to gain an intuitive 
understanding why the CKM matrices have the qualitative features they 
do and to show that they emerge from the basic structure of the scheme 
as simple consequences of classical differential geometry and can be 
deduced, almost quantitively in some cases, without a detailed calculation 
\cite{features}.

\begin{figure}[htb]
\vspace{-5cm}
\centerline{\psfig{figure=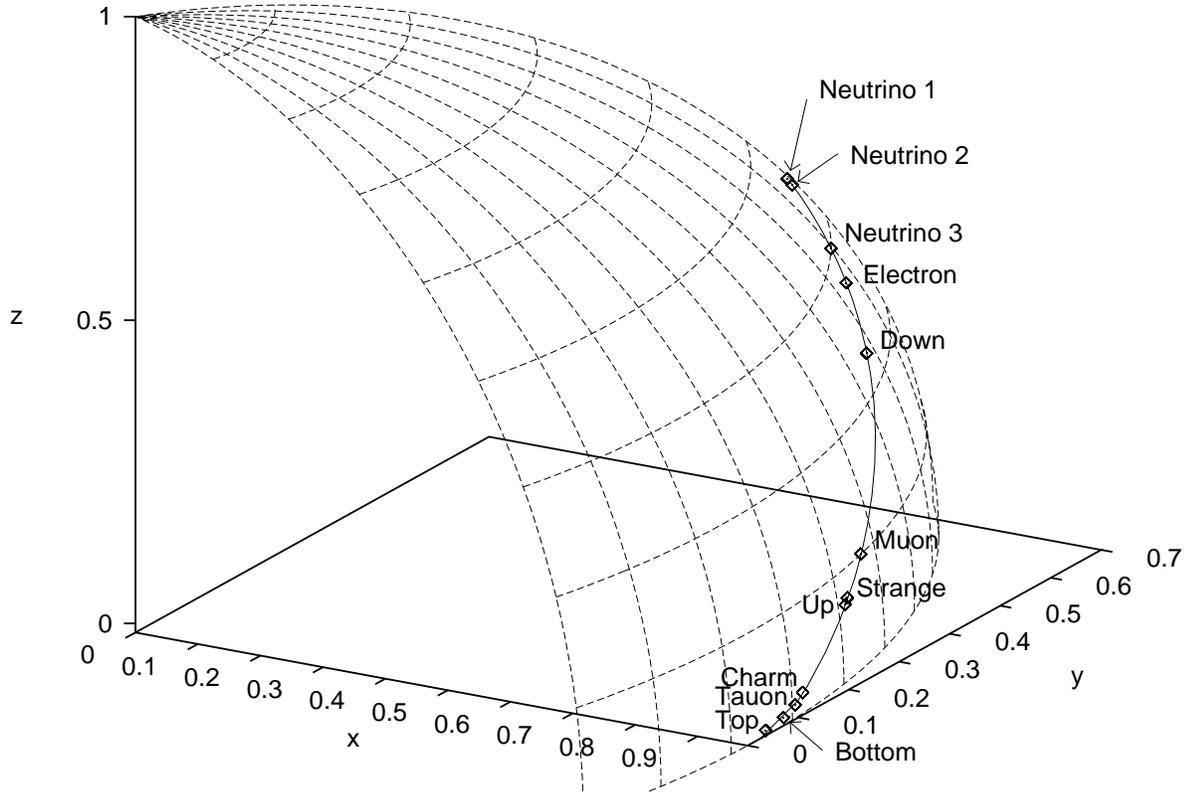,width=0.85\textwidth}}
\vspace{0cm} 
\caption{The trajectory traced out by $(x',y',z')$ and the locations on
it of the 12 fermion states.}
\label{runtraj}
\end{figure}

We recall that the fermion mass matrix in the DSM scheme is factorized
even after loop corrections, and all the information needed for our
consideration of the CKM matrix is encoded in the normalized vector 
$(x',y',z')$.  This  vector rotates with the energy scale, tracing
out a trajectory on the unit sphere, which trajectory is the same
for all fermion types $T$, i.e. whether $U$- or $D$-type quarks, charged
leptons or neutrinos.  The various physical states, however, differ in 
the locations they occupy on this trajectory.  Figure \ref{runtraj} 
shows the actual trajectory and locations of the 12 fermion states 
obtained in the fit of \cite{ourckm,ournuos}.

\begin{figure}[htb]
\centerline{\psfig{figure=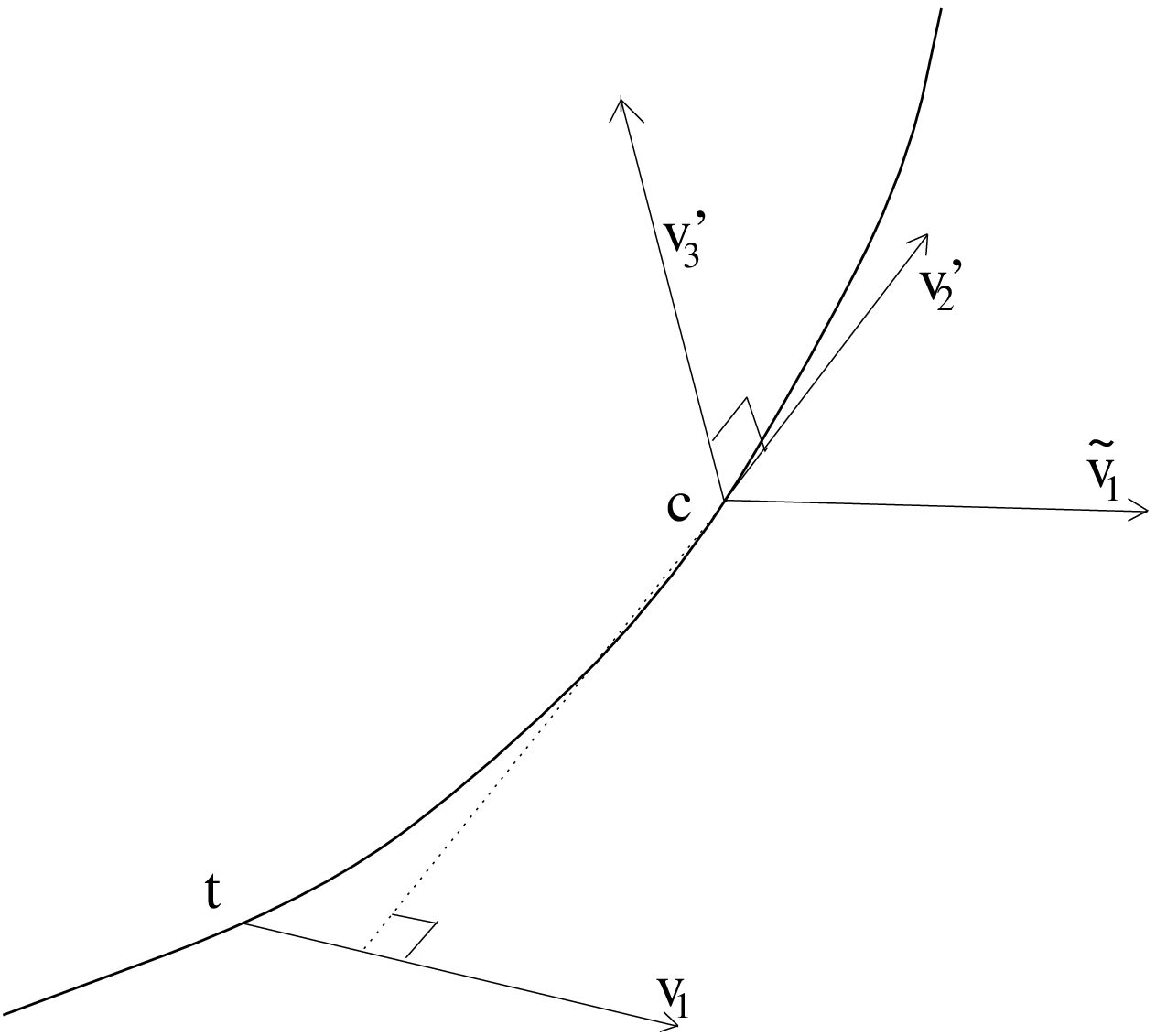,width=0.6\textwidth}}
\vspace{0cm} 
\caption{The state vectors of the 3 physical states belonging to the
3 generations of the $U$-type quark.}
\label{v1v2v3}
\end{figure}

The state vectors of the various physical states are given in terms of
the rotating vector $(x',y',z')$ as follows.  (i) Evaluated at the scale 
of the top mass, $(x',y',z')$ is the state vector $|{\bf v}_1 \rangle$ 
of $t$, as shown in Figure \ref{v1v2v3}.  (ii) At the scale of the charm 
mass, the vector $(x',y',z')$ is rotated to another direction, say 
$|{\bf \tilde{v}}_1 \rangle$ in Figure \ref{v1v2v3}, with thus a zonzero 
component orthogonal to $|{\bf v}_1 \rangle$, the direction of which 
gives the state vector $|{\bf v}'_2 \rangle$ of $c$.  (iii) The state 
vector of the $u$-quark is ${\bf v}'_3 = {\bf v}_1 \wedge 
{\bf v}'_2$.
Similar constructions apply to the other 3 fermion types $D, L, N$.

\begin{figure}[htb]
\centerline{\psfig{figure=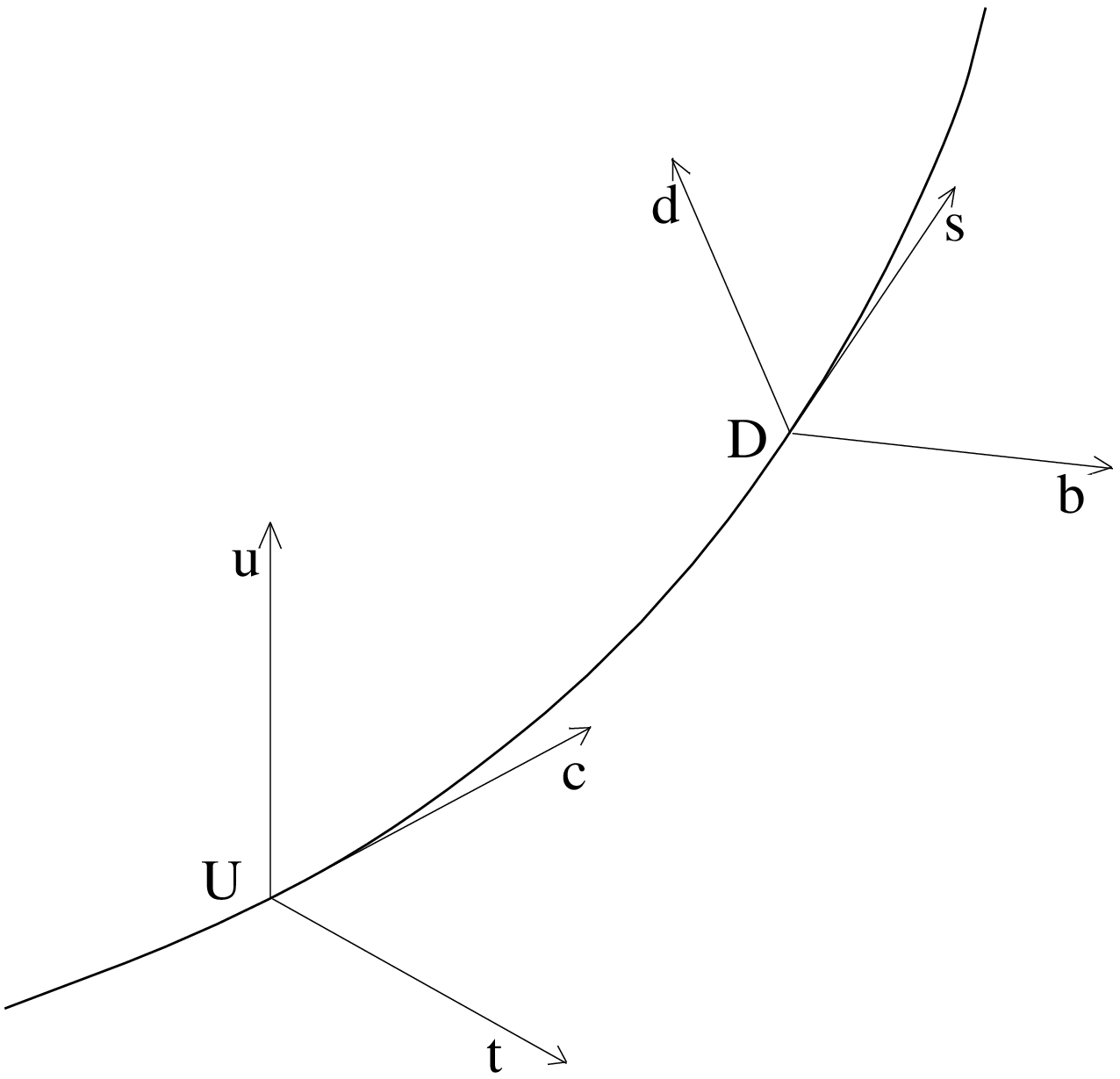,width=0.6\textwidth}}
\vspace{0cm} 
\caption{Two triads of state vectors for two fermion types
transported along a common trajectory.}
\label{2triads}
\end{figure}

If we make the approximation that the locations of the $t$ and $c$ quarks 
on the trajectory are close together (as is seen to be true in Figure 
\ref{runtraj}), then the 3 state vectors of $t,c,u$ of Figure \ref{v1v2v3} 
form in that limit an orthonormal triad at the $t$ position.  If we do 
the same for the $D$-type quarks, we have another such triad at the $b$ 
position, as illustrated in Figure \ref{2triads}.  The entries of the 
(quark) CKM matrix are nothing but the direction cosines between the 
vectors of these 2 triads.  The leptonic CKM matrix is similar.

Since the trajectory lies on the unit sphere, the $c$ vector is the tangent 
{\bf T} to the trajectory and the $t$ vector the normal {\bf N} to the surface,
so that they form, together with the $u$ vector ${\bf B}={\bf N} \wedge
{\bf T}$, what is known in elementary differential geometry as the 
`Darboux triad'.  Differentiating then with respect to the arc-length, 
we get the following formulae similar to the well-known Serret--Frenet 
formulae for space curves \cite{docarmo}:
\begin{eqnarray}
{\bf N}' & = & -\kappa_n {\bf T} - \tau_g {\bf B}, \nonumber\\
{\bf T}' & = &  \kappa_g {\bf B} + \kappa_n {\bf N}, \nonumber \\
{\bf B}' & = & -\tau_g {\bf N} - \kappa_g {\bf T}.
\label{SFDarboux}
\end{eqnarray}
Here $\kappa_g$ is the geodesic curvature, $\kappa_n$ the normal curvature, 
and $\tau_g$ the geodesic torsion of the curve on the surface.  Equivalently, 
to first order in arc-length $\Delta s$, (\ref{SFDarboux}) can be rewritten 
in the form of a CKM matrix with entries arranged in the conventional order:
\begin{equation}
\left( \begin{array}{ccc}
       1 & -\kappa_g \Delta s & -\tau_g \Delta s \\
       \kappa_g \Delta s & 1 & \kappa_n \Delta s \\
       \tau_g \Delta s & -\kappa_n \Delta s & 1 \end{array} \right).
\label{ckmdg}
\end{equation}

In our case of the unit sphere, $\tau_g=0$ and $\kappa_n=1$.  It follows 
then from (\ref{ckmdg}) that \cite{features}: 
\begin{description}
\item{(a)} The corner elements of both the quark CKM matrix ($V_{ub}, 
V_{td}$) and the leptonic CKM matrix ($U_{e3}, U_{\tau 1}$) are small 
since they vanish to first order in the separation between the
corresponding fermion types.
\item{(b)} The 4 other off-diagonal elements of the quark CKM matrix
are small compared to the diagonal elements since they are of first 
order in the separation between the $t$ and $b$ quarks, which is small
as seen in Figure~\ref{runtraj}.
\item{(c)} The elements $V_{cb}, V_{ts}$ for quarks are much smaller
than their counterparts $U_{\mu 3}, U_{\tau 2}$ for leptons, since they
are to first order proportional to the separation, which is much smaller
for quarks than for leptons as seen in Figure~\ref{runtraj}.
\end{description}
These 3 points are all borne out by experiment as already noted above.
Indeed, it is amusing to note that even the approximate values for the 
4 elements in (c), as quoted above from either experiment (\ref{expckmq}),
(\ref{expckml}) or the DSM calculation (\ref{calckmq}), (\ref{calckml}),
can simply be read off by measuring the separations between $t$ and $b$ 
and between $\tau$ and $\mu$ on the trajectory in Figure~\ref{runtraj}!

Further, we note that since the geodesic curvature $\kappa_g$, in contrast 
to the geodesic torsion $\tau_g$ and the normal curvature $\kappa_n$ on 
a sphere, depends both on the location and on the trajectory, so do the 
values of the remaining pair of off-diagonal elements of the CKM matrix, 
namely the `Cabibbo angles' ($V_{us}, V_{cd}$ for quarks, and $U_{e2}, 
U_{\mu 1}$ for leptons).  This explains why the Cabbibo angle is so large 
even though the separation between $t$ and $b$ is small.  It also means 
that the 12 elements are much more sensitive to the details of the fit 
and explains why our calculation has been less successful in predicting
$U_{e2}$ than with the other leptonic mixing angles.

That all these features in the mixing matrices echoing experimental data 
can be derived {\em without} detailed calculations is very encouraging, 
for it means that the agreement with experiment reported in the 2 sections
before are much less likely to be just numerical accidents of the calculation.

\setcounter{equation}{0}

\section{FCNC Effects from Dual Gluon Exchange}

Besides explaining the features of the Standard Model, any scheme which 
attempts to go beyond has of course also to examine its own predictions 
for the possibility of their violating already some known experimental 
limits, and if not, for the feasibility of their being tested by future 
experiment.  For the DSM scheme, one obvious direction to probe in 
this respect is the new interactions arising from exchanges of the dual 
colour gauge bosons.  Dual colours in DSM having been identified with 
generations, it follows that any particle carrying a generation index 
can acquire a new interaction by exchanging these bosons, leading to 
generation-changing or flavour-changing neutral current (FCNC) effects.  
These gauge bosons are presumably quite heavy or otherwise they would 
already have been discovered.  There are thus two areas where one can 
look for their influence.  One can either look for effects at energies 
low compared with their mass where the effects of their exchange would 
be suppressed, or else for sizeable effects at ultra-high energies.  We 
shall consider examples of both, at low energies in this section, and 
at high energies in the next.

At low energy, flavour-changing neutral currents can manifest themselves
in rare decays and in mass differences between charge conjugate neutral 
meson pairs.  For DSM, as for other `horizontal gauge symmetry' models
\cite{buchmuller}, these effects arise already at the level of one-(FCNC) 
gauge boson exchange and can thus be estimated once given the masses of 
the gauge bosons and their couplings to the fermions involved.  What 
distinguishes DSM, however, is that the scheme has been made so restrictive 
by what has gone before that detailed estimates can now be given for all 
these FCNC effects at the one-boson exchange level in terms of only one 
additional parameter. 
 
In view both of the intrinsic structure built into the scheme and of the 
calculations already performed which are summarized above, most of the
`fundamental' parameters of DSM as at present formulated are now known.  
First, by virtue of the Dirac quantization condition \cite{dualcomm}:
\begin{equation}
g_3 \tilde{g}_3 = 4\pi, \;\;\; g_2 \tilde{g}_2 = 4\pi, \;\;\;
g_1 \tilde{g}_1 = 2\pi,
\label{couplings}
\end{equation}
the coupling strengths $\tilde{g}_i$ of the dual gauge bosons are derivable 
from the coupling strengths $g_i$ of the ordinary colour and electroweak 
gauge bosons measured in present experiments.  Secondly, the branching 
of these couplings $\tilde{g}_i$ into the various physical fermion states 
are given by the rotation matrices relating these physical states to the
`gauge states' in generation or dual colour space, thus:
\begin{equation}
\psi_{gauge,L}^T = S^T \psi_{physical,L}^T
\label{rotationf}
\end{equation} 
where the index $T$ runs over the four types of fermions $U, D, L$ and
$N$.  These matrices $S^T$ were already determined in the calculation 
of fermion mixing matrices \cite{ourckm,ournuos}, where for example 
the (quark) CKM matrix was obtained as $(S^U)^\dagger S^D$
in Section 2 and there found to be in excellent agreement with experiment.  
Finally, in tree-level approximation, the masses of the dual gauge bosons 
are given in terms of the vacuum expectation values of the dual colour 
Higgs fields, the ratios $x, y, z$ between which are among the parameters 
determined in the calculation \cite{ourckm} by fitting the CKM matrix.  Thus 
the only remaining unknown among the quantities required is the actual 
strength $\zeta$ of the vev's, which, though also entering in principle
in the calculations of Standard Model parameters outlined in Section 2, 
turns out to be hardly restricted there.  That being the case, one can 
now calculate in the DSM scheme all one-dual gauge boson exchange diagrams 
between any two fermions in terms of this single mass parameter $\zeta$.  

There are altogether 9 dual gauge bosons (including that corresponding
to $\tilde{U}(1)$) which can be exchanged, whose masses at tree-level
are given by diagonalizing a mass matrix dependent on the dual couplings
$\tilde{g}_3, \tilde{g}_1$ and on $\zeta$ and $(x, y, z)$.  Given that,
as mentioned in Section 2, the value obtained for $(x, y, z)$ from fitting 
the quark CKM matrix \cite{ourckm} is very close to the fixed point 
$(1,0,0)$, the mass matrix for the dual gauge bosons can be readily 
diagonalized \cite{ourfcnc} yielding one particular state with mass:
\begin{equation}
M^2 = \zeta^2 z^2  \frac{3}{4} \frac{\tilde{g}_3^2}
{1 + \frac{3}{16}\frac{\tilde{g}_3^2}{\tilde{g}_1^2}},
\label{gmass}
\end{equation}
which is much lower than the rest.  As a result, the calculation for FCNC 
effects becomes quite simple, being dominated by the exchange of just 
this one boson.  

At energies much lower than the mass of this dual gauge boson, the effects
can then be summarized in terms of an effective Lagrangian thus \cite{ourfcnc}:
\begin{equation}
L_{eff}=
\frac{1}{2 \zeta^2 z^2} 
\sum_{T,T'} f^{T,T'}_{\alpha,\beta;\alpha',\beta'} 
(J^{\mu \dagger}_T)^{\alpha,\beta}  (J_{\mu,T'})^{\alpha',\beta'}\,,
\label{laeff1}
\end{equation}
with currents of the usual $\;V-A\;$ form:
\begin{equation}
(J^{T}_\mu)_{\alpha,\beta} = \bar{\psi}_{L,\alpha}^T \gamma_\mu 
\psi_{L,\beta}^T,
\label{current2}
\end{equation}
and a group factor which, for reactions involving changes of flavour, 
reduces to:
\begin{equation}
f^{T,T'}_{\alpha,\beta;\alpha',\beta'} =
S^{T*}_{3,\alpha} S^{T}_{3,\beta}
S^{T'*}_{3,\alpha'} S^{T'}_{3,\beta'},
\label{groupfactor1}
\end{equation}
which is given entirely in terms of the matrices $S^T_{\alpha,\beta}$,
so that the only remaining free parameter in (\ref{laeff1}) is the mass 
scale $\zeta z$.  

However, the effective Lagrangian (\ref{laeff1}) describes only interactions 
between quarks and leptons.  To make contact with actual experiment on 
hadrons, we follow the usual procedures adopted in these contexts.  
For example, the effective action gives a contribution to the $K_L-K_S$ 
mass difference of the form:
\begin{equation}
\Delta m_K =
\frac{1}{\zeta^2 \,\, z^2} |f^{D,D}_{2,3;2,3}|
\langle K^0| \left[\bar{s}_L \gamma^\mu d_L \right]^2 |\bar{K}^0 \rangle.
\end{equation}
Evaluating the matrix elements in the vacuum insertion approximation one 
obtains:
\begin{equation}
\Delta m_K =
\frac{1}{\zeta^2 \,\, z^2} |f^{D,D}_{2,3;2,3}|\frac{f_K^2 \, \, m_K}{3},
\label{delmK}
\end{equation}
where $f_K$ is the $K$ decay constant and $m_K$ is the $K$ mass.  Mass
differences between other charge conjugate neutral mesons are treated
similarly.  On the other hand, for hadron decays, in order to minimize 
the uncertainties in the hadron structure we take quotients between the 
rare and Standard Model-allowed processes which contain the same or similar
hadronic matrix elements.  For instance, for $K^+$ decays we take:
\begin{equation}
\frac
{Br \left(K^+ \rightarrow \pi^+ l_{\alpha} l_{\beta} \right)}
{Br \left(K^+ \rightarrow \pi^0 \nu_{\mu} \mu^+ \right)}
=  |f^{D,L}_{2,3;\alpha,\beta}|^2 
\left( \frac{v}{\zeta \, \, z} \right)^4 \frac{2}{\sin^2 \theta_c},
\label{kplus}
\end{equation}
where $v=\frac{0.246}{\sqrt{2}}$ TeV and $\theta_c$ is the Cabibbo angle, 
$\sin \theta_c=0.23$.  Similarly, for the leptonic decays of the neutral 
$K$-mesons, we take:
\begin{equation}
\frac {\Gamma \left(K^0_{L(S)} \rightarrow l_\alpha l_\beta \right)}
   {\Gamma \left(K^+ \rightarrow \mu^+ \nu_\mu \right)}
= |f^{D,L}_{2,3;\alpha,\beta}|^2
  \left( \frac{v}{\zeta \, \, z} \right)^4 \frac{1}{\sin^2 \theta_c}.
\label{kls}
\end{equation}
Then from these formulae, given the total widths of the $K$'s and their 
widths in the Standard Model-allowed modes as measured in experiment, one 
can easily calculate the branching ratios of the various rare modes of 
$K$-decay.  These procedures for dealing with the complexities of hadronic
effects are of course far from foolproof but are likely to give the rough
order of magnitudes correctly.

All predictions we obtain in this way are still dependent on the single
parameter $\zeta z$, so that without any further input we can give no
numerical value for the predicted quantities.  So long as $\zeta z$ 
remains undetermined, our predictions will of course lead to no violation
of present experiment.  However, given the experimental bound on any 
one quantity, a bound on $\zeta z$ is implied, which will then allow us 
to give the correlated bound on all the others.  The most stringent 
lower bound on $\zeta z$ obtained in this way turns out to be that from 
the experimentally measured $K_L-K_S$ mass difference, namely $\Delta 
m_K(K_L-K_S) =3.5 \,\times \, 10^{-12}\mbox{ MeV}$ which is of roughly
the right order of magnitude expected from second order weak interactions.
Requiring that the FCNC effect due to dual gauge boson exchange be smaller
than this value gives the bound \cite{ourfcnc}:
\begin{equation}
\zeta \,\, z \geq 400\mbox{ TeV}.
\label{bound}
\end{equation}
The correlated (upper) bounds on other FCNC effects due to dual gauge 
boson exchange can then be estimated.  

As will be seen in the next section, there is a possible upper bound on 
the parameter $\zeta z$ coming from a rather unexpected angle which turns 
out to be similar to the lower bound quoted in (\ref{bound}).  If that is
the case, then the above bounds for FCNC effects can be treated as actual 
order of magnitude estimates.

For $\zeta z =$ 400 TeV, the predicted branching ratios \cite{ourfcnc} of some 
rare $K$-decay modes are given in Table \ref{Kdecaytab} and compared with 
the experimental limits/measurements \cite{databook}.  One notes that most 
of the predictions are way beyond the present experimental sensitivity, 
while some others, such as $K_L \rightarrow e^+ e^-, \mu^+ \mu^-$, can also
go by second order weak interactions which are expected to give similar
or even somewhat larger contributions and so will overshadow the present
predicted effects.  Only the mode $K_L \rightarrow e^\pm \mu^\mp$, which
is inaccessible to second order weak interactions unless neutrinos mix,
is interesting in having a predicted branching ratio less than two orders 
of magnitude down from the present experimental limits \cite{brookhaven}
and so may be accessible in the near future.  Its observation at this level
may be considered as a confirmation either of the DSM prediction or that
neutrinos mix and hence of interest in either case.

\begin{table}
\begin{eqnarray*}
\begin{array}{||l|l|l||}
\hline \hline
  & Theory & Experiment \\
\hline
Br(K^+ \rightarrow \pi^+ e^+ e^-)  &  4 \times 10^{-15}   &
2.7 \times 10^{-7}  \\
Br(K^+ \rightarrow \pi^+ \mu^+ \mu^-)  & 2 \times 10^{-15}    &
2.3 \times 10^{-7}  \\
Br(K^+ \rightarrow \pi^+ e^+ \mu^-)  &  2 \times 10^{-15}   &
7 \times 10^{-9}  \\
Br(K^+ \rightarrow \pi^+ e^- \mu^+)  &  2 \times 10^{-15}  &
2.1 \times 10^{-10}   \\
Br(K^+ \rightarrow \pi^+ \nu {\bar \nu})  &  2 \times 10^{-14}   &
2.4 \times 10^{-9}  \\
Br(K_L \rightarrow e^+ e^-)  & 2 \times 10^{-13}   &
4.1 \times 10^{-11}  \\
Br(K_L \rightarrow \mu^+ \mu^-)  & 7 \times 10^{-14}    &
7.2 \times 10^{-9} \\
Br(K_L \rightarrow  e^{\pm} \mu^{\mp})  &  1 \times 10^{-13}    &
5.1 \times 10^{-12} \\
Br(K_S \rightarrow \mu^+ \mu^-)  &  1 \times 10^{-16}   &
3.2 \times 10^{-7}\\
Br(K_S \rightarrow  e^+ e^-)  &  3 \times 10^{-16}   &
2.8 \times 10^{-6} \\
\hline \hline
\end{array}
\end{eqnarray*}
\caption{Branching ratios for rare leptonic and semileptonic $K$ decays.  
The first column shows the DSM predictions from one-dual gauge boson exchange
with the lowest v.e.v. $\zeta z$ of the Higgs fields taken at 400 TeV.  The
second column gives either the present experimental limits on that process
if not yet observed or the actual measured value for that process.  In the
latter case, it means that the process can go by other mechanisms such as 
second-order weak so that our predictions with dual gauge boson exchange
will appear as corrections to these.  Except for the entry for the decay
$K_L \rightarrow e \mu$ from \cite{brookhaven} mentioned in the text, the 
other entries are from the databook \cite{databook}.}
\label{Kdecaytab}
\end{table}

Similar tables have been compiled for rare $D$ and $B$ meson decays but 
the predicted branching ratios are in all cases much below the present
experimental sensitivity and therefore not of immediate interest.

Mass differences between the conjugate neutral $D$ and $B$ meson pairs 
are given in a similar way to that for the $K$'s.  The contribution of 
one-dual gauge boson exchange to the mass splitting in $D$ is \cite{ourfcnc}:
\begin{equation}
{\Delta m_D} = \frac{m_D}{\zeta^2 z^2} \frac{f_D^2}{3} |f^{U,U}_{2,3;2,3}|.
\label{Dmassdiff}
\end{equation}
Taking the values $f_D^2=10^{-8}\mbox{ TeV}^2$ for the $D$-meson coupling and
$m_D=1.865$ GeV for the mass we have:
\begin{equation}
{\Delta m_D} = 5 \times 10^{-12}\mbox{ MeV}.
\label{Dmassnos}
\end{equation}
This is one-and-a-half orders of magnitude off the present experimental 
limits ${\Delta m_D}\leq 1.4 \times 10^{-10}$ MeV and could be accessible 
to planned experiments in the near future.  Applying the same procedure 
to the mass-splitting between the neutral $B$-mesons, one finds that the 
contribution from dual gauge boson exchange is 6 orders of magnitude 
smaller than that from the Standard Model and thus not likely to be 
accessible.

We conclude therefore, for this section, that the DSM predictions on 
low energy FCNC effects so far made do not seem in contradiction with 
any existing experiment, and that for a couple of cases, namely 
$K_L \rightarrow e^\pm \mu^\mp$ and $\Delta m(D-\bar{D})$, apart of course
from $\Delta m(K_L-K_S)$ itself, they may be testable in the near future.

\setcounter{equation}{0}

\section{Air Showers beyond the GZK Cut-off}

At energies higher than the mass scale of the dual gauge bosons, FCNC 
effects will no longer be suppressed and, given the strength of their 
couplings $\tilde{g_i}$ as governed by the Dirac quantization conditions 
(\ref{couplings}), the interactions due to their exchange will become 
strong.  Hence, DSM would predict a new strong interaction at ultra-high
energies for all particles carrying a generation index.  In particular,
even the neutrinos corresponding to the usual 3 generations of charged
leptons will acquire strong interactions at high energies.  At first
sight, this seems alarming until one recalls from the estimate given in
the last section for the mass scale involved of order $\zeta z > 
400\ {\rm TeV}$, 
which is way beyond anything that has been achieved in terrestial experiments
or can be achieved in the foreseeable future.  Nor are such energies
accessible to astrophysical or cosmological considerations  except in the 
very early universe.  There is in fact only one instance known to us that 
energies of that order have been experimentally observed, namely in air showers
produced by cosmic rays with energy beyond the Greisen-Zatsepin-Kuz'min 
(GZK) cutoff.

Air showers with energy $E > 10^{20}\ {\rm eV}$ \cite{Volcano}--\cite{Auger}, 
though rare in occurrence, pose a long-standing and intriguing question of 
fundamental physical interest.  High energy air showers are usually thought
to be due to protons, but protons with such an energy would interact with 
the 2.7 K microwave background via, for example, the reaction:
\begin{equation}
p + \gamma_{2.7 K} = \Delta + \pi,
\label{piprod}
\end{equation}
and degrade quickly in energy.  Indeed, according to Greisen, Zatsepin 
and Kuz'min \cite{Greisen,Zatsemin}, the cosmic ray spectrum for protons 
should cut off sharply at around $5 \times 10^{19}\ {\rm eV}$ (the GZK 
cut-off) if they come from further than 50 Mpc away.  And within such 
distances, there does not seem to be any likely source for producing 
particles of so high an energy.

One possible solution would be that these post-GZK air showers are produced 
not by protons but by some other (stable, electrically neutral) particles 
which would not be so degraded in energy by the microwave background.  The 
possibility can thus be considered that they are produced by neutrinos, which 
is feasible, of course, only if for some reason neutrinos acquire at high 
energy a new strong interaction, for otherwise they would not interact
sufficiently with air nuclei to produce air showers.  But this is exactly 
what is predicted by DSM as proposed in the preceding paragraph.  So, it would 
appear that this prediction not only may escape contradiction with existing 
experiment as one might at first have feared, but may even offer an 
explanation for the long-standing puzzle of air showers beyond the GZK
cutoff \cite{airshower1,airshower2,airshower3}.

However, a strong interaction for neutrinos, though necessary, is by 
itself insufficient to guarantee a large cross section with air nuclei, 
which is needed for them to produce air showers in the atmosphere, since 
whatever the strength of the interaction, the cross section will remain 
small if the interaction is short-ranged.  Now, the dual gauge bosons 
in DSM being supposedly very heavy, it looks at first sight that the 
interactions they mediate will be short-ranged and therefore not lead 
to large high energy cross sections for neutrinos.  But this need not 
be the case for the following reason.  The relation of dual colour to 
colour is similar to that of magnetism to electricity in electrodynamics 
which has only the physical degrees of freedom corresponding to the 
one photon, despite having two separate, electric {\it and} magnetic, 
gauge symmetries, as explained in Section 1.  Hence, dual colour, though 
a different gauge degree of freedom to colour, represents just the same 
physical degree of freedom as colour \cite{dualsymm}.  As a result, it 
is argued \cite{airshower2} that a dual gluon can `metamorphose' into 
a gluon in hadronic matter, thus giving neutrinos at high energy an
interaction of hadronic range inside the nucleus.  They will then interact 
coherently with the air nucleus and acquire with it a hadronic size cross 
section.\footnote{There has appeared a paper by Burdman, Halzen and 
Gandhi \cite{Halzen} subsequent to \cite{airshower2} claiming, among other 
things, that neutrinos cannot on general grounds acquire hadronic size cross 
sections.  We think that their claim is ill-founded.  Their arguments used 
only first order perturbation theory which is far from adequate for hadron 
reactions which are notoriously nonperturbative in character.  Indeed, with 
their arguments, one would be unable to deduce that protons have hadronic 
cross sections.  They also claimed their conclusions to be a consequence
of s-wave
unitarity but gave neither justifications nor references to substantiate this 
claim, and a repeated effort by one of us (CHM) in correspondence with Francis
Halzen has not produced any clarification.  We do not think s-wave unitarity
can constrain the cross section the way they claim it does since high
energy cross sections involve all partial waves.  For more details of 
this discussion see \cite{airshower3,ourfcnc}.}  This assertion is admittedly
rather conjectural.  If it fails, then the suggested explanation for 
post-GZK showers no longer stands, but the assertion that the prediction 
of strong neutrino interactions at high energy contradicts no existing 
experiment still remains valid.

Since this suggested explanation for post-GZK air showers depends 
crucially on the assumed identification of generations with dual colour, 
it is worth examining its feasibility in some detail.
Suppose that neutrinos do acquire strong interactions and a large enough 
cross section with air nuclei to produce air showers at high energy.  The 
energy at which this begins to happen, according to DSM, is given by the 
scale estimated before to be $>$ 400 TeV in the centre of mass.  For a
neutrino impinging on a nucleon at rest in the atmosphere, this corresponds
to a primary energy of around $8 \times 10^{19}\ \rm{eV}$, namely just above
the GZK cut-off, exactly the sort of energy one wants.  That being the case,
let us now examine in more detail whether the hypothesis can accommodate 
the few observed facts known about the post-GZK showers, most of which 
have difficulties in being explained by protons as the primary particle.  

(A) First, one asks how neutrinos at such a high energy can be produced.  
One does not actually know at present a truly realistic mechanism even for
protons, but according to Hillas \cite{Hillas} one can at least write down 
the condition that a source must satisfy in order to produce such energetic 
particles:
\begin{equation}
BR > E/Z,
\label{Hillas}
\end{equation}
where $B$ is the magnetic field in $\mu$G, $R$ the size of the source 
in kpc, $E$ the energy in EeV = $10^{18}\ {\rm eV}$ and $Z$ the 
charge of the particle.  There are only a few types of objects known
which satisfy this condition, namely neutron stars, radio galaxies and
active galactic nuclei.  Of these, both the neutron stars and active 
galactic nuclei are surrounded by strong electromagnetic fields.  The
difficulty with protons is that even if the source can accelerate them
to the required high energy, they would not be able to escape from the
intense fields surrounding the source.  However, there does not seem to
be the same difficulty with neutrinos.  By hypothesis based on the DSM,
neutrinos interact strongly at high energy so that any source capable
of accelerating protons to these energies will be able also copiously 
to produce neutrinos by say proton-proton collisions.  Once produced,
however, neutrinos will not be deterred by the intense e.m. fields and
will be able to escape where protons fail.

(B) Secondly, one asks whether neutrinos will be able to survive a long
journey through the 2.7 K microwave background.  There is no problem, 
for in colliding with a (massless) photon at this temperature, a neutrino
even at $10^{20}\ {\rm eV}$ will produce only about 200 MeV C.M. energy,
and at this energy a neutrino has still only weak interactions.  The same
applies also to its collision with the neutrino background in the intervening 
space.

(C) Thirdly, one asks whether a neutrino when it arrives on earth will 
be 
able to produce air showers with the observed angular and depth distributions.
Neutrinos with only weak interactions will have immense penetrating power, 
and even if an enormous neutrino flux is assumed sufficient to produce air 
showers at these rare occasions, the showers will be mostly horizontal 
and have a near constant distribution in depth.  This is in contradiction 
to what is observed for post-GZK showers which are mostly near vertical and
occur in the upper atmosphere.  However, once neutrinos are ascribed a
hadronic cross section with air nuclei, they will interact like hadrons
and both the angular and depth distributions will automatically fall into 
place.

(D) It was noted \cite{Hayashida} that out of the few post-GZK
events seen, three pairs coincide in incident angles to within the
experimental error of about $2^\circ$.  The probability of this occurring
at random is very small and the obvious conclusion would be that the 
two members of each pair originate from the same source.  They have
however different energies and if they are protons should be deflected
differently by the intervening magnetic fields and hence arrive at
different angles, contrary to what is observed.  If they are neutrinos,
on the other hand, they will not be deflected by magnetic fields and
will arrive on earth in the same direction they started out.

(E) The highest energy event at $3 \times 10^{20}\ {\rm eV}$ observed
by the Fly's Eye \cite{Flyseye} was noted to point in the direction of a 
very powerful Seyfert galaxy 900 Mpc away \cite{Elmers}.  If that is 
taken to be its source and if it is due to a proton, one would wonder why
many more showers with lower energies are not observed from the same
direction, for a powerful source capable of accelerating protons to such
a high energy would surely also produce protons at lower energies as well.
For neutrinos interacting strongly only at high energy, this is not a 
problem.  At low energies, neutrinos are weakly interacting and would
first of all not be produced at source, and even if produced would not
give rise to air showers when they arrive on earth.

It thus seems that the neutrino hypothesis has survived all the above tests 
(A)--(E) on post-GZK showers which pose difficulties for their 
having been produced by protons.  In spite of this, however, the hypothesis 
must clearly be subjected to many more quantitative tests before it can be 
taken seriuosly.  Fortunately, some such tests \cite{airshower2,ourfcnc}
are available, as follows.

(I) If we accept our previous argument that a neutrino at high energy 
would interact not only strongly but coherently with an air nucleus, it is easy
to deduce from a geometric picture that the cross section of a neutrino
with the nucleus would be about half that of a proton.  To both the
neutrino and the proton, the nucleus would appear as a black disc, say of
radius $r_A$.  The neutrino, with as yet no known internal structure,
would appear to the nucleus still as a point, but the proton will appear
again as a black disc of radius, say, $r_p$.  One concludes thus that
the (geometric) cross section of the neutrino with the nucleus is roughly:
\begin{equation}
\sigma_T(\nu A) = \pi r_A^2,
\label{sigmanu}
\end{equation}
while that of the proton with the nucleus is roughly:
\begin{equation}
\sigma_T(p A) = \pi (r_p + r_A)^2.
\label{sigmap}
\end{equation}
Assuming that:
\begin{equation}
r_A \sim A^{1/3} r_p,
\label{rA}
\end{equation}
with $A$ around 15 for an air nucleus, one easily deduces the above estimate:
\begin{equation}
\frac{\sigma_T(\nu A)}{\sigma_T(p A)} \sim \frac{1}{2}.
\label{sigmaratio}
\end{equation}
This means that neutrinos at post-GZK energies are expected to be about
twice as penetrating as protons, and hence that neutrino-produced air showers
will occur at a lower depth on the average than proton-produced air
showers.  Folding in the air density as a function of height, it is easy
to evaluate the penetration probability as a function of depth.  This
was done in \cite{airshower2} which finds that:
\begin{eqnarray}
{\rm Most \; probable \; height \; of} \; p{\rm -produced \; showers} 
   & \sim & {\rm 21 \; km}, \nonumber \\
{\rm Most \; probable \; height \; of} \; \nu{\rm -produced \; showers} 
   & \sim & {\rm 15 \; km}.
\label{showerdepth}
\end{eqnarray}
Hence, one predicts that post-GZK air showers which are supposedly produced 
by neutrinos would occur most probably at a height of only around 15 km,
in contrast to lower energy showers produced by protons which would occur 
most probably at a height of around 21 km in our atmosphere.  Present 
detectors do not locate the primary vertices of air showers readily.  
For this reason, we have found up to the present only one tentative piece 
of information for testing this prediction.  The development profile of the 
highest energy event obtained by the Fly's Eye shows that light began to be 
observed at a (vertical equivalent) height of around 12 km.  If we interpret
this as the primary vertex for the event, then it is much more likely,
according to the preceding arguments, to be a neutrino-produced shower than 
a proton-produced one, for which the probability is estimated to be 
less than 5 \%.  This conclusion should not as yet be taken too seriously,
but with new projects such as Auger \cite{Auger}, capable of collecting
sizeable statistics, this could be a very useful test for the hypothesis
that post-GZK showers are neutrino-produced.

(II) As far as particle physics proper is concerned, the post-GZK 
air shower events, if interpreted as due to neutrinos, are useful in 
providing a rough upper bound to the dual gauge boson mass.  Translated 
to the language of Section 5, this means an upper bound on the parameter 
$\zeta z$ of around 500 TeV \cite{ourfcnc}, remarkably close to the 
lower bound of around 400 TeV obtained there from the $K_L-K_S$ 
mass difference.  Acceptance of this upper bound then converts the bounds 
estimated in Section 5 on rare meson decays and mass differences into 
actual order of magnitude predictions, and hence affords a second test 
for the neutrino hypothesis here for post-GZK air showers.

\setcounter{equation}{0}

\section{Concluding Remarks}

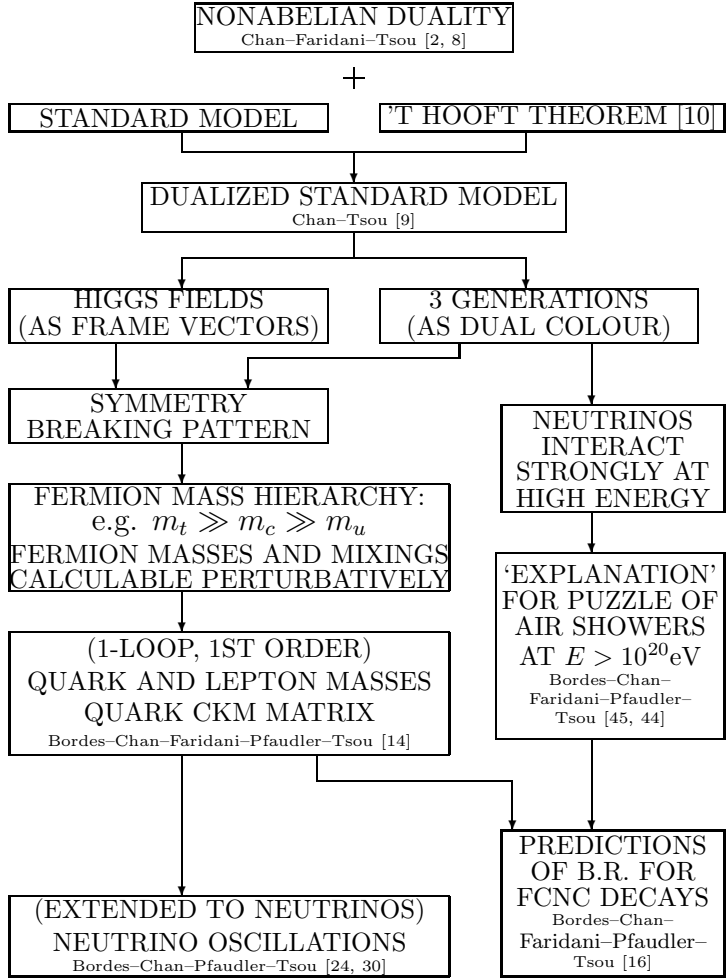
\begin{figure}[htb]
\center
\begin{picture}(240,400)
\put(60,350){\framebox(120,18){\shortstack{{\footnotesize NONABELIAN DUALITY}\\
{\tiny{Chan--Faridani--Tsou \cite{dualsymm,dualsym}}}}}}
\put(116,340){\line(1,0){8}}
\put(120,344){\line(0,-1){8}}
\put(-10,320){\framebox(120,10){{\footnotesize STANDARD MODEL}}}
\put(130,320){\framebox(130,10){{\footnotesize 'T HOOFT THEOREM
\cite{tHooft}}}}
\put(55,320){\line(0,-1){8}}
\put(185,320){\line(0,-1){8}}
\put(55,312){\line(1,0){130}}
\put(120,312){\vector(0,-1){12}}
\put(40,282){\framebox(160,18){\shortstack{{\footnotesize DUALIZED
STANDARD MODEL}\\
{\tiny{Chan--Tsou \cite{dualcons}}}}}}
\put(120,282){\line(0,-1){10}}
\put(55,272){\line(1,0){130}}
\put(55,272){\vector(0,-1){12}}
\put(185,272){\vector(0,-1){12}}
\put(-10,240){\framebox(120,20){\shortstack{{\footnotesize HIGGS FIELDS}\\
{\footnotesize (AS FRAME VECTORS)}}}}
\put(130,240){\framebox(120,20){\shortstack{{\footnotesize 3 GENERATIONS}\\
{\footnotesize (AS DUAL COLOUR)}}}}
\put(30,240){\vector(0,-1){18}}
\put(160,240){\line(0,-1){6}}
\put(80,234){\line(1,0){80}}
\put(80,234){\vector(0,-1){12}}
\put(-10,202){\framebox(120,20){\shortstack{{\footnotesize SYMMETRY}\\
{\footnotesize BREAKING PATTERN}}}}
\put(55,202){\vector(0,-1){16}}
\put(-10,146){\framebox(166,40){\shortstack{{\footnotesize FERMION MASS 
HIERARCHY:}\\
e.g. $m_t \gg m_c \gg m_u$\\{\footnotesize FERMION MASSES AND MIXINGS}\\
{\footnotesize CALCULABLE PERTURBATIVELY}}}}
\put(55,146){\vector(0,-1){16}}
\put(-10,84){\framebox(166,46){\shortstack{{\footnotesize (1-LOOP, 1ST
ORDER)}\\
{\footnotesize QUARK AND LEPTON MASSES}\\{\footnotesize QUARK CKM MATRIX}\\
{\tiny{Bordes--Chan--Faridani--Pfaudler--Tsou
\cite{ourckm}}}}}}
\put(106,84){\line(0,-1){10}}
\put(106,74){\line(1,0){74}}
\put(180,74){\vector(0,-1){19}}
\put(55,84){\vector(0,-1){54}}
\put(-10,0){\framebox(166,30){\shortstack{{\footnotesize (EXTENDED TO 
NEUTRINOS)}\\
{\footnotesize NEUTRINO OSCILLATIONS}\\
{\tiny{Bordes--Chan--Pfaudler--Tsou
\cite{ournuos,features}}}}}}
\put(210,240){\vector(0,-1){24}}
\put(176,176){\framebox(84,40){\shortstack{{\footnotesize NEUTRINOS}\\
{\footnotesize INTERACT}\\{\footnotesize STRONGLY AT}\\ 
{\footnotesize HIGH ENERGY}}}}
\put(210,176){\vector(0,-1){16}}
\put(174,90){\framebox(86,70){\shortstack{{\footnotesize `EXPLANATION'}\\
{\footnotesize FOR PUZZLE OF}\\{\footnotesize AIR SHOWERS}\\ 
{\footnotesize AT
$E>10^{20}$eV}\\
{\tiny{Bordes--Chan--}}\\{\tiny{Faridani--Pfaudler--}}\\
{\tiny{Tsou \cite{airshower3,airshower2}}}}}}
\put(210,90){\vector(0,-1){35}}
\put(176,0){\framebox(84,55){\shortstack{{\footnotesize PREDICTIONS}\\
{\footnotesize OF B.R.\ FOR}\\{\footnotesize FCNC DECAYS}\\
{\tiny{Bordes--Chan--}}\\{\scriptsize{Faridani--Pfaudler--}}\\
{\tiny{Tsou \cite{ourfcnc}}}}}}
\end{picture}
\caption{Summary flow-chart}
\label{flowchart}
\end{figure}

The basic tenets and applications to-date of the DSM scheme are summarized
in the flow chart of Figure \ref{flowchart}, which shows that starting
from a previously derived result of nonabelian duality, one is led on the
one hand to a calculation of some of the Standard Model's fundamental
parameters, and on the other to new testable predictions ranging from 
FCNC effects at low energy to air showers from cosmic rays at the extreme 
end of the detected energy scale.

One of the most attractive features of DSM is undoubtedly its offer of a 
possible explanation for the existence both of exactly 3 fermion generations 
and of scalar Higgs fields.  In the conventional formulation of the Standard 
Model, the necessity to introduce by hand both of these, neither having any 
known geometrical significance, must be regarded as rather a blotch on a gauge
theory otherwise so beautifully founded on geometry.  The identification thus 
of generations as dual colour and of Higgs fields as frame vectors in 
internal symmetry space, giving each a geometrical significance, seems very 
attractive.  Besides, according to \cite{dualsymm}, nonabelian duality is 
an intrinsic property of the Standard Model (as of any gauge theory) which 
then brings with it automatically a 3-fold broken dual colour symmetry 
and frame vectors in internal symmetry space playing a dynamical 
role.  In other words, the niches for 3 fermion generations and Higgs fields 
already `pre-exist' in the Standard Model.  Hence, it seems appropriate to 
assign them to just these features we see in nature, because even if we 
do not, we shall still have to account for them physically in some other 
way.

Implementing these identifications with some seemingly natural assumptions 
as detailed in Section 1, one is then led to a scheme with a hierarchical
fermion mass spectrum where the mixings between fermion types and lower 
generation masses are calculable as loop corrections in terms of a few
parameters.  The present score from the 1-loop calculation carried out 
to-date is as follows.  By adjusting 3 parameters, namely the 
(common) Yukawa coupling strength $\rho$ and the 2 ratios between the
3 vev's $(x,y,z)$ of the dual colour Higgs fields, one has calculated 
the following 14 among the 26 or so of the Standard Model's fundamental
parameters: the 3 independent parameters in the quark CKM matrix $|V_{rs}|$, 
the 3 corresponding parameters in the leptonic CKM matrix $|U_{rs}|$, and 
the 8 masses $m_c, m_s, m_\mu, m_u, m_d, m_e, m_{\nu_1}, B$.  Of these 14 
calculated quantities, 2 (i.e.\ $m_u$, $U_{e2}$) compare unsatisfactorily 
at present with experiment, and another 2 (i.e. the mass of the lightest 
neutrino $m_{\nu_1}$ and that of the right-handed neutrino $B$) are untested 
being experimentally yet unknown.  The other 10, however, (namely, $|V_{rs}|, 
|U_{e3}|, |U_{\mu3}|, m_c, m_s, m_\mu, m_d, m_e$), all agree as well as 
can be expected with their known empirical values.  This, we think, is not 
a bad score for a first attempt based on some rather crude approximations,
such as taking $\rho$ and $m_T$ as scale-independent constants. With more 
experience and sophistication, the score can possibly be improved.

However, even if considered successful, this score by itself does not 
constitute a stringent test for the basic assumption of DSM that dual colour 
is generations.  As emphasized in a recent paper \cite{phenodsm}, the same 
result can be obtained just by assuming generations to be a broken $U(3)$
symmetry independent of whether it is identified with dual colour.  The 
only physical consequence considered in this paper which relies crucially
on that identification is the explanation suggested in Section 6 for
air showers beyond the GZK cut-off, which is still far from established.  
An urgent task for this scheme is thus to device some further tests for 
the dual colour hypothesis.   

Besides this, there are many further questions needing answers for checking 
the consistency of DSM, both within itself and with nature.  Of these we list 
in particular the following.  First, there is the question of CP-violation
which, though known experimentally, has not yet made an appearance in DSM 
at the 1-loop level and might indicate a deficiency.  Secondly, there is 
the question of the rotating mass matrix, which one has made use of in 
the calculation of mass and mixing parameters, and this might have other 
physical consequences yet waiting to be explored.  Thirdly, there is 
the intriguing question of the `accidental' near equality between Yukawa 
coupling strengths $\rho$ for all fermion types, and the proximity of the 
Higgs vev's $(x, y, z)$ to the fixed point $(1,0,0)$, which presumably 
reflect a deeper intricacy in the problem than we have yet understood.
Fourthly, there is the question of linkage between the breaking of the 
dual colour symmetry, studied so far in isolation, with the breaking 
of the electroweak symmetry, which may be related to the point raised
immediately above.  Fifthly, there is the subtle question of whether the 
duality assertion that gauge and dual gauge bosons represent the same 
physical degrees of freedom might give rise to a new class of phenomena,
called `metamorphosis' by us in \cite{dualcons}, of which post-GZK air 
showers are but one example of many possible manifestations.  Sixthly, 
going even further afield, there is the question of the symmetry
$\widetilde{SU}(2)$ dual to electroweak $SU(2)$, which by the same 
logic adopted for $SU(3)$ colour here, ought to give rise to another 
level of confinement deeper than colour, and this should be amenable to 
experimental investigation, but only by deep inelastic scattering at 
ultra-high energies.  And there will be other questions too which we 
have not yet learned even to formulate. One has thus the feeling that 
what has been attempted so far is but scratching the surface of a 
possibly very rich vein.

\vspace{.5cm}

\noindent {\large {\bf Acknowledgement}}

\vspace{.2cm}

It is a pleasure for us to thank Jos\'e Bordes, Jacqueline Faridani and 
Jakov Pfaudler for a most enjoyable and fruitful collaboration producing
most of the work reported above.


\end{document}